\documentclass[aps,pre,twocolumn,superscriptaddress,10pt]{revtex4-2}
\usepackage{silence}

\usepackage{amsmath,amssymb,graphicx,bm,physics}
\usepackage[utf8]{inputenc}
\usepackage{fix-cm}
\usepackage{newtxtext,newtxmath}
\usepackage{color}
\usepackage[american]{babel}
\usepackage{caption}
\usepackage{subcaption}
\DeclareCaptionJustification{fulljustified}{\leftskip=0pt\rightskip=0pt\parfillskip=0pt plus 1fil\relax}
\usepackage[hidelinks]{hyperref}

\makeatletter
\chardef\l@en=\l@american
\makeatother

\begin{document}

\title{{Gap-controlled thermalization in a chain of alternating nonlinear springs: an SSH model version of the Fermi-Pasta-Ulam-Tsingou problem}}

\author{José A. Aké}
\affiliation{Posgrado en Ciencias F\'{i}sicas, Universidad Nacional Autónoma de México, Ciudad de México, Mexico}
\author{Gerardo G. Naumis}
\affiliation{Depto. de Sistemas Complejos, Instituto de F\'{i}sica, Universidad Nacional Aut\'{o}noma de M\'{e}xico (UNAM). Apdo. Postal 20-364, 01000 M\'{e}xico D.F., M\'{e}xico}

\date{\today}

\begin{abstract}
In classical anharmonic lattices, the resonance structure driving nonlinear mode mixing is reshaped by band gaps. However, it remains unclear whether a spectral gap hinders or promotes long-time thermalization. Here we study a Fermi-Pasta-Ulam-Tsingou (FPUT) chain of equal masses joined by alternating dimensionless spring constants  $1\pm\Delta\kappa$, so that the chain is dimerized with strength $\Delta\kappa$ --- the classical vibrational analogue of the Su–Schrieffer–Heeger (SSH) model --- and quantify how the acoustic-optical gap controls relaxation under $\alpha$-type (cubic) nonlinearity. Tracking modal energies and spectral entropy via long-time symplectic simulations, we find that long time relaxation is delayed by the dimerization. When the dimerization strength reaches \(\Delta\kappa=1/2\), set by the onset of the first umklapp process that allows two zone-boundary acoustic phonons to fuse into a zone-center optical phonon;  the three-wave acoustic–acoustic–optical channel is closed. By studying  the effective acoustic-to-
optical scattering rate, it is found that there is an interplay between the dynamic resonance vertex and the kinematic resonance manifold, which evolve differently as a function of the dimerization parameter affecting thermalization. Above the  threshold \(\Delta\kappa=1/2\), energy redistribution can nevertheless continue through higher-order resonances and quasiresonances. Yet specific boundary conditions and initial conditions can inhibit equipartition. These results identify the phononic gap as a tunable, momentum-selective filter for nonlinear energy transport, suggesting a general route to controlling energy transfer and equipartition in dimerized or topologically gapped nonlinear lattices and in flexible mechanical systems.
\end{abstract}

\maketitle

\section{Introduction}
The Fermi-Pasta-Ulam-Tsingou (FPUT) problem was originally conceived to explore the routes to thermal equilibrium in classical nonlinear systems \cite{fermi1955studies}. Contrary to expectations from statistical mechanics, FPUT chains displayed long-lived recurrences and a lack of thermalization for specific initial conditions \cite{zabusky1965interaction}. These findings sparked decades of research in nonlinear dynamics, soliton theory, and chaos  \cite{ford1992fermi,berman2005fermi,campbell2005introduction,gallavotti2007fermi}.

Recent interest has grown in generalizations of the FPUT system that incorporate spatial heterogeneity or topological features \cite{verhulst2020variations, many2022nonlinear,wang2024thermalization, lin_fu_wang_zhang_zhao_2025,pezzi_multi-wave_2025}.  Previous studies have focused on chains of particles with alternating masses \cite{pezzi_multi-wave_2025}; the present work addresses a chain of particles with alternating spring stiffnesses, i.e., a parametrically controlled dimerized chain in which one type of spring is progressively softened. This model is the classical vibrational analogue of the celebrated Su–Schrieffer–Heeger (SSH) model in solid-state physics, in which a tight-binding chain has two alternating transfer integrals. Its importance stems from the fact that it is the simplest model of a topological Hamiltonian. Moreover, previous studies have focused on homogeneous or weakly modulated FPUT lattices, leaving the influence of a robust band gap on long-time energy redistribution largely unexplored. The present study systematically investigates how SSH-like dimerization \cite{asboth2016short} controls interbranch energy transfer through a tunable band gap.

While in monoatomic cubic chains six-wave resonances typically dominate the nonlinear transfer \cite{OnoratoPNAS_2015} , the chain with two different masses plus a cubic term $\alpha$-chain admits three-wave interactions (two acoustic + one optical) that can produce faster local activation of the optical band \cite{onorato2023wave, pezzi_multi-wave_2025}. However, these triad resonances are often isolated and therefore insufficient for complete thermalization \cite{bustamante2019exact}, so four-wave (and higher-order) processes act on longer timescales to produce global equipartition \cite{onorato2023wave, pezzi_multi-wave_2025} for the case of the system with two different masses.

\begin{figure}[!ht]
    \centering
    \includegraphics[width=1.0\linewidth]{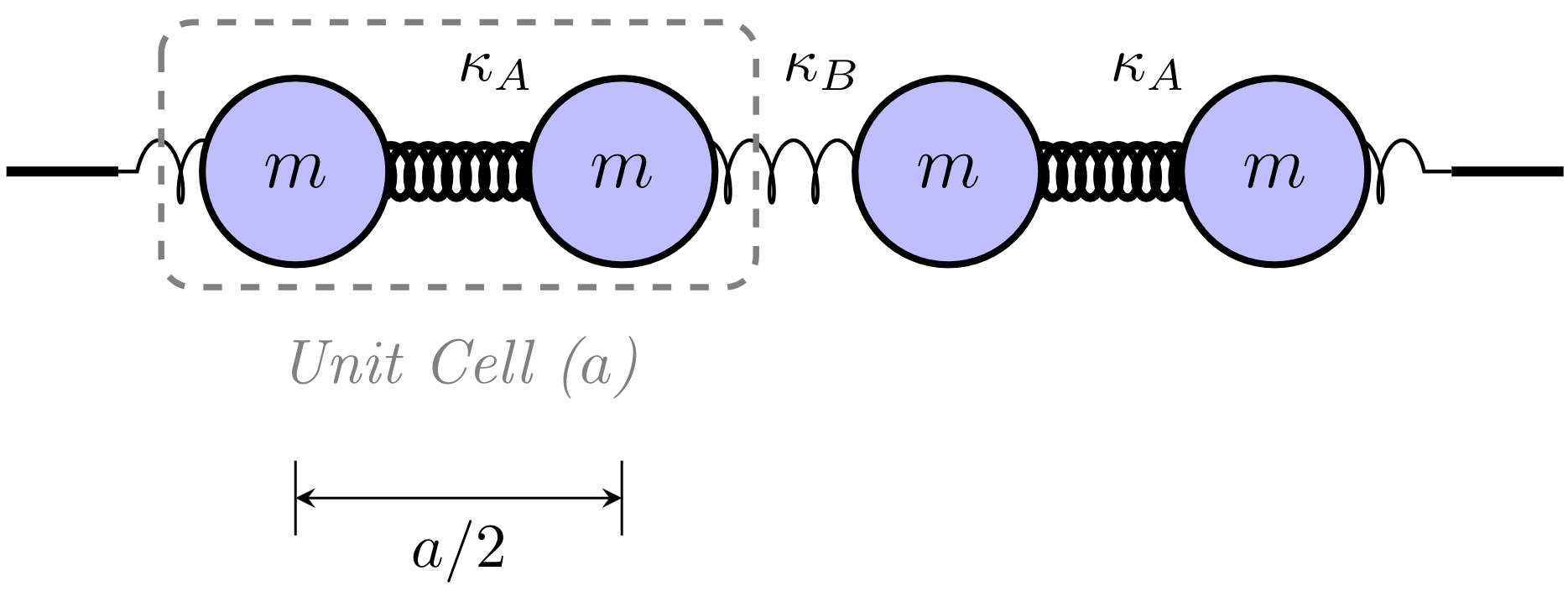}
    \caption{Schematic of the alternating FPUT chain with spring constants $\kappa_{A} = 1+\Delta\kappa$ and $\kappa_{B} = 1-\Delta\kappa$ connecting identical masses. As $\kappa_B \rightarrow 0$, the system is dimerized in pairs of masses joined by the non linear spring with stiffness $\kappa_A$. For $\Delta\kappa=1$, the optical branch corresponds to the oscillations of such $N/2$ dimers, while the other $N/2$ modes become the center of mass movement of each dimer, with zero frequency  oscillations. Thus the system becomes flexible in the sense of rigidity theory as the acoustic branch is made by $N/2$ floppy modes.} 
    \label{fig:diatomic_chain_diagram}
\end{figure}

A relevant open question is whether dimerized spring constants in this classical SSH analogue model can delay thermalization once finite-amplitude nonlinearities are included. This question arises from classical mechanics in weak-nonlinearity theory: low-order resonant triads (two acoustic + one optical) can efficiently bridge the gap when frequencies satisfy a resonance condition, yet for sufficiently large gap width $\Delta \omega$, these resonances vanish and interbranch energy transfer can be suppressed. Thus, the interplay between gap width and nonlinearity becomes the critical control parameter. Previous work on systems with different masses has been systematically studied by Galgani et. al\cite{galgani1992problem} and Maiocchi\cite{maiocchi2019freezing} from an analytical and numerical perspective and giving an answer for these class of systems. Nevertheless, the case of the system with different spring constants, despite proving valuable insights remains mainly unexplored. Although both systems possess acoustic and optical branches, the parameter controlling the band separation and the nonlinear coupling is physically different. Rather than establishing a general long-time perturbative bound for the energy of the optical modes \cite{maiocchi2019freezing}, here  we identify the mechanism responsible for the slowdown of acoustic--optical energy transfer. In particular, our analysis of the nonlinear interaction coefficients and three-wave resonance conditions shows that, for a certain threshold, no three-wave resonance connects the acoustic and optical branches. The long-lived low spectral entropy observed numerically is consequently interpreted as delayed, rather than blocked, interbranch energy transfer and incomplete equipartition on the simulated time scales. Recent work by Manda et al. \cite{many2022nonlinear, pal2018amplitude} and Sone et al. \cite{sone2025} investigated nonlinear edge-states delocalization in SSH-like mechanical lattices, exploiting the same chiral bulk-boundary topology that classifies our dimerized chain \cite{su1979solitons, heeger1988solitons, susstrunk2016classification, huber2016topological, coutant2021acoustic}; our study complements this bulk-boundary picture by instead probing acoustic-mode spectral transfer within the bulk, and how the phononic gap controls the density and accessibility of resonant triads.

In this work, we study FPUT chains with alternating spring constants  with focus on the $\alpha$-type (cubic) nonlinearity. Simulations are performed under fixed boundary conditions (FBC) and periodic boundary conditions (PBC), starting from either single-mode or localized initial excitations. Therefore, we explore the thermalization of the phononic equivalent of the SSH model. We analyze modal energies and spectral entropy $S(t)$, following the methodology in \cite{wang2024thermalization, lin_fu_wang_zhang_zhao_2025}.  
The layout of this paper is as follows. Section \ref{sec:model_and_methods} introduces the FPUT chain model with alternating spring constants, the linear dispersion relation and gap structure, the nonlinear Hamiltonian and couplings in normal modes, observables, entropy and details in numerical simulations.  Section \ref{sec:results} reports the main numerical results. Section \ref{sec:discussion} presents the three-wave resonance conditions, derives the threshold $\Delta\kappa=1/2$ and interprets these results in terms of the resonant manifold, the strength of the nonlinear couplings and their relation via the effective scattering rate, and discusses the physical implications. Section \ref{sec:conclusions} outlines future perspectives. More details on calculations can be found in section \ref{sec:appendix}.

\section{Model and Methods}\label{sec:model_and_methods}

This section presents the  chain made of  equal masses joined by alternating nonlinear springs  (dimerized FPUT)  model and the analytical and computational methods employed to characterize its thermalization dynamics. We first introduce the physical model and its linear dispersion properties  (\ref{subsec:physical_model}), then describe the normal mode transformation (\ref{subsec:normal_modes}), define the observables used to quantify thermalization (\ref{subsec:observables}), and finally detail the numerical implementation (\ref{subsec:numerics}).

\subsection{Physical model}\label{subsec:physical_model}

We considered a one-dimensional FPUT chain consisting of $N$ particles of equal masses (dimensionless units $m = 1$) connected by alternating spring constants (Fig.~\ref{fig:diatomic_chain_diagram}). Spring $j$ joins particles $j$ and $j+1$. Linear stiffnesses alternate: \(\kappa_j=\kappa_A\) for odd \(j\), \(\kappa_j=\kappa_B\) for even \(j\). 
In dimensionless units, the stiffnesses can also be written as, 
\begin{equation}
    \kappa_j = 1 - (-1)^{j} \Delta\kappa
\end{equation}
with $\kappa_A=1+\Delta\kappa$ and $\kappa_B=1-\Delta\kappa$ and $\Delta \kappa$ is here called the dimerization strength. Nonlinear coefficient stiffnesses can also alternate: \(\gamma_j=\gamma_A\) for odd \(j\), \(\gamma_j=\gamma_B\) for even \(j\), where $\gamma_{A} = \kappa_{A}\alpha$ and $\gamma_{B} = \kappa_{B}\alpha$ (so that $\gamma_B/\gamma_A=\eta=\kappa_B/\kappa_A$).

The Hamiltonian of the system (up to cubic order in the potential) is,
\begin{equation}
    H=\sum_{j=1}^N \frac{p_j^2}{2m}
  +\sum_{j}\left[\frac{1}{2}\kappa_j\,(q_{j+1}-q_j)^2+\frac{1}{3}\gamma_j\,(q_{j+1}-q_j)^3\right].
  \label{eq:FPU_Hamiltonian}
\end{equation} 
For periodic boundary conditions $q_{N+1}\equiv q_1$ and the spring sum runs over $j=1,\dots,N$ (the closing spring $j=N$ is of type B). For fixed boundary conditions $q_0=q_{N+1}=0$ and the sum runs over $j=0,\dots,N$, i.e.\ it includes the two springs attached to the walls (for even $N$ both are of type B); omitting the spring $j=0$ would leave the first particle free on one side, in contradiction with the equations of motion below. Notice that unlike the standard FPUT models where the nonlinearity was uniform, here the nonlinear coefficients are locally scaled by the spring constant $\kappa_{j}$, ie., $\gamma_A=\kappa_A \alpha$ for odd and $\gamma_B=\kappa_B \alpha$, where $\alpha$ is the strength of the third-order terms potentials. 
This choice is physically motivated by the fact that linear and nonlinear elastic constants emerge from the same interatomic potential. Hamilton's equations give $m\ddot q_j = f_j - f_{j-1}$, with bond force $f_j = \kappa_j[(q_{j+1}-q_j) + \alpha(q_{j+1}-q_j)^2]$.

\subsection{Linear Hamiltonian case}

Applying Bloch's theorem to the linearized equations of motion yields the dispersion relation for alternating spring constants $\kappa_A$ and $\kappa_B$. The Bloch construction below is exact for PBC; for FBC see the remark at the end of this subsection.) Introduce a two-site unit cell indexed by \(n=1,\dots,M\) with \(M=N/2\).
Label sublattice components as
\[
q_{n,1}=q_{2n-1},\qquad q_{n,2}=q_{2n}.
\]
Spring $2n-1$ (type A) joins $q_{n,1}$ and $q_{n,2}$, and spring $2n$ (type B) joins $q_{n,2}$ and $q_{n+1,1}$.
Define Bloch transforms (unit-cell normalization) using units of $a$ such that $a=1$ in what follows,
\[
q_{n,l}=\frac{1}{\sqrt{M}}\sum_k Q_{l}(k)\,e^{ikn},\qquad
p_{n,l}=\frac{1}{\sqrt{M}}\sum_k P_{l}(k)\,e^{ikn},
\]
where \(k=2\pi \ell/M,\ \ell=0,\dots,M-1\) (the index $l=1,2$ labels the sublattice). The linearized equations are
$m\ddot q_{n,1}=-(\kappa_A+\kappa_B)q_{n,1}+\kappa_A q_{n,2}+\kappa_B q_{n-1,2}$ and
$m\ddot q_{n,2}=-(\kappa_A+\kappa_B)q_{n,2}+\kappa_A q_{n,1}+\kappa_B q_{n+1,1}$. By proposing a solution that oscillates in time with frequency $\omega(k)$ and using the Bloch transforms, this leads to the eigenvalue equation,
\begin{equation}
\mathbf{D}(k)\mathbf{e}_{k,\sigma}
=m\omega^{2}_\sigma(k)\mathbf{e}_{k,\sigma},
\qquad \sigma=\pm,
\end{equation}
where the \(2\times2\) dynamical matrix (for the linear problem) is,
\begin{equation}
\mathbf{D}(k)=
\begin{pmatrix}
\kappa_A+\kappa_B & -\mathcal{S}(k)\\
-\mathcal{S}^*(k) & \kappa_A+\kappa_B
\end{pmatrix},
\end{equation}
where we define,
\[
\mathcal{S}(k)=\kappa_A+\kappa_B e^{-ik},\qquad |\mathcal{S}(k)|=\sqrt{\kappa_A^2+\kappa_{B}^2+2\kappa_A\kappa_B\cos k}.
\]
and which yields eigenvalues,
\[
m\omega_{\pm}^2(k)=(\kappa_A+\kappa_B)\pm|\mathcal{S}(k)|.
\]
A convenient normalized choice of eigenvectors \(\boldsymbol{e}_{k,\sigma}\), where \(\sigma=+,-\) labels the optical and acoustic branches, is

\begin{equation}
\boldsymbol{e}_{k,-}
=
\frac{1}{\sqrt{2}}
\begin{pmatrix}
s(k)\\
1
\end{pmatrix},
\qquad
\boldsymbol{e}_{k,+}
=
\frac{1}{\sqrt{2}}
\begin{pmatrix}
-s(k)\\
1
\end{pmatrix},
\end{equation}

with

\begin{equation}
s(k)=\frac{\mathcal S(k)}{|\mathcal S(k)|}\equiv e^{i\phi(k)},
\qquad
\zeta(k)=e^{ik}s(k),
\end{equation}

which satisfy \(\boldsymbol{e}_{k,\sigma}^\dagger \boldsymbol{e}_{k,\sigma'}=\delta_{\sigma\sigma'}\). Note that $s(-k)=s^*(k)$, hence $\boldsymbol e_{-k,\sigma}=\boldsymbol e_{k,\sigma}^{*}$, $s(0)=1$ and, for $\Delta\kappa>0$, $s(\pi)=\mathrm{sgn}(\kappa_A-\kappa_B)=+1$. Project modal coordinates onto these eigenvectors,
\[
Q_{k,\sigma}=\sum_{l=1}^2 e_{k,\sigma}^{(l)*}\,Q_l(k),\qquad
P_{k,\sigma}=\sum_{l=1}^2 e_{k,\sigma}^{(l)*}\,P_l(k),
\]
where $e_{k,\sigma}^{(l)}$  for $l=1,2$ is the $l$ component of the vector $\boldsymbol{e}_{k,\sigma}$. The inverse transformation is $Q_l(k)=\sum_\sigma e^{(l)}_{k,\sigma}Q_{k,\sigma}$, and for real displacements $Q_{-k,\sigma}=Q_{k,\sigma}^{*}$ (same branch index $\sigma$).
The linear (quadratic) Hamiltonian in modal coordinates is diagonal,
\[
H_{\mathrm{lin}}=\sum_{k,\sigma}\left(\frac{|P_{k,\sigma}|^2}{2m}+\frac{1}{2}m\omega_{\sigma}^2(k)\,|Q_{k,\sigma}|^2\right).
\]

\textit{Remark (FBC).} For fixed ends the translational symmetry is broken and the linear modes are the eigenvectors of the $N\times N$ stiffness matrix; they are standing waves, i.e.\ superpositions of the Bloch states $\pm k$, with the same two bands but a different quantization of $k$ and boundary corrections of order $1/N$. All FBC modal energies $E_{k,\sigma}$ below are defined in that eigenbasis, and the Bloch-based resonance analysis applies to FBC up to such corrections.

Let us now analyze in more detail the dispersion. It is explicitly given by,
\begin{equation}
    \omega_{\pm}^{2}(k) = \frac{\kappa_{A} + \kappa_{B}}{m} \pm \frac{1}{m}\sqrt{(\kappa_{A}+\kappa_{B})^{2} - 4\kappa_{A}\kappa_{B} \sin^{2}{(k/2)}},
    \label{eq:dispersion_relation_different_springs}
\end{equation}

\begin{figure}[!ht]
    \centering
    \includegraphics[width=1.02\linewidth]{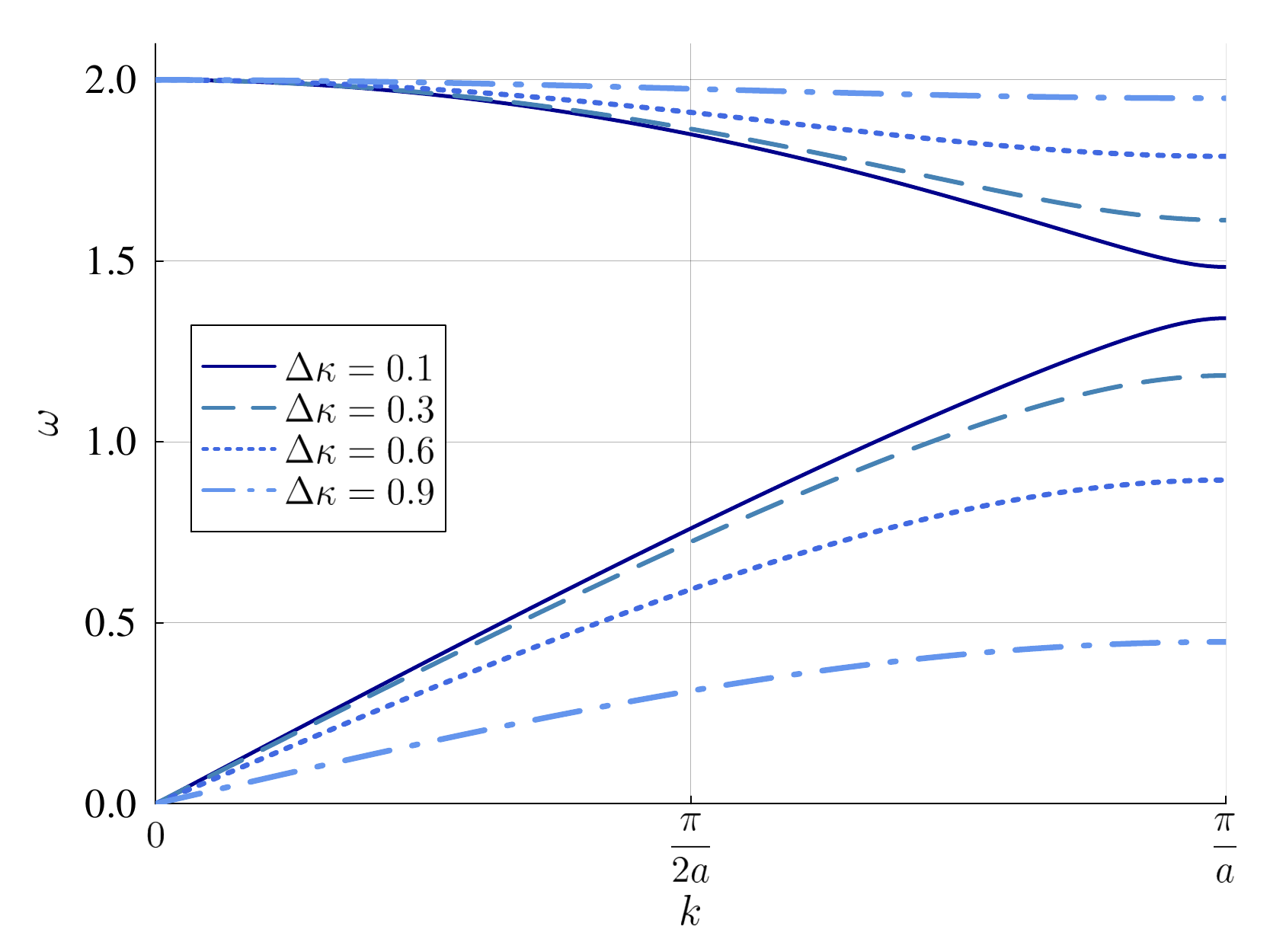}
    \caption{Dispersion relation [Eq.~(\ref{eq:dispersion_relation_different_springs})] for varying $\Delta\kappa$, showing the frequency gap at $k = \pi/a$ plotted in the reduced zone scheme. The gap grows as $\Delta \kappa$ grows, setting the frequency barrier that must be bridged by nonlinear resonances for interbranch energy transfer to occur.}
    \label{fig:dispersion_relation}
\end{figure}

where $a$ is the lattice constant, $k \in [-\pi/a, \pi/a]$ the wavevector in the first Brillouin zone, and the $\pm$ signs corresponded to the optical and acoustic branches, respectively. This can be rewritten as, 
\begin{equation}
    \omega^{2}_{\pm}(k) = \frac{2\bar{\kappa}}{m} \left(1 \pm \sqrt{1 - \kappa^{*} \sin^{2}{(k/2)}}\right),
    \label{eq:dispersion_relation_different_springs_bonita}
\end{equation}
where,
\begin{equation}
    \bar{\kappa}=(\kappa_{A} + \kappa_{B})/2=\frac{\kappa_{A}}{2}(1+\eta),
\end{equation} is the average spring constant and,
\begin{equation}
    \kappa^{*}=\frac{\kappa_{A}\kappa_{B}}{\bar{\kappa}^2}=\frac{4\eta}{(1+\eta)^{2}}
\end{equation}
is a dimensionless spring-contrast parameter ($0<\kappa^*\le1$, with $\kappa^*=1$ for the uniform chain), where we also defined the spring elastic constants ratio,
\begin{equation}
\eta=\kappa_B/\kappa_A,
\end{equation}

If  $\omega^{2}(k)$ is measured in dimensions of $\bar{\kappa}/m$, for simplification we can use units where $m=1$. By assuming that $\kappa_{A} = 1 + \Delta\kappa$ and $\kappa_{B} = 1 - \Delta\kappa$, then
\begin{equation}
    \kappa^{*}= 1-\Delta\kappa^2, \quad \bar{\kappa}=1
    \label{eq:reduced_dispersion_relation}
\end{equation}
and $\eta$ is now defined as 
\begin{equation}
    \eta = \frac{1-\Delta \kappa}{1+\Delta \kappa}
\end{equation}

Figure~\ref{fig:dispersion_relation} illustrates the energy dispersion for different $\Delta \kappa$ values as obtained from Eq. (\ref{eq:dispersion_relation_different_springs}). At the Brillouin zone boundary ($k = \pi$), a frequency gap of width $\Delta\omega$ emerges that separates the two branches (Fig.~\ref{fig:dispersion_relation}). Notice how the frequency gap increases with $\Delta\kappa$ and progressively isolates the two branches. The acoustic branch spans frequencies from $0$ to $\omega^{max}_{\mathrm{-}} = \sqrt{2(1-\Delta\kappa)}$ while the optical branch extends from $\omega^{min}_{\mathrm{+}} = \sqrt{2(1+\Delta\kappa)}$ to $2$.  In the limit $k \to 0$, the dispersion relation for the acoustic branch simplifies to
\begin{equation}
    \omega^{2}_{-} (k) \approx v_{s}^{2}k^{2}
\end{equation}
where
\begin{equation}    
    v_{s} = \sqrt{\frac{\kappa^{*}}{4}}
    \label{eq:speed_of_sound_diatomic_chain}
\end{equation}
is the speed of sound of the system (in units of cells per unit time). Two interesting limiting cases arise here: (i) the uniform chain, recovered 
by setting $\Delta\kappa=0$, and (ii) the limit of weakly coupled strong 
springs, obtained for $\Delta\kappa \rightarrow 1$. The ii) limiting case implies that $\kappa^{*} \rightarrow 0$ and then, 
\begin{equation}\label{eq:dispersion_relation_limit_floppy}
    \omega^{2}_{-}(k) \approx  \kappa^{*} \sin^{2}{(k/2)},
\end{equation}
 Notice that as $\kappa^{\ast} \to 0$ the entire acoustic branch flattens to $\omega_{-} \approx 0$ for all $k$ in Eq. \ref{eq:dispersion_relation_limit_floppy}. These $N/2$ modes have nearly zero frequency (Fig. \ref{fig:diatomic_chain_diagram}). In the limit $\Delta \kappa \to 1$, the optical branch approaches the vibration frequency of isolated pairs of masses joined by a strong spring, i.e., dimers. These dimers are weakly coupled by the weak springs, resulting in a kind of dimer liquid.
The acoustic branch becomes an $N/2$-fold degenerate mode, corresponding to the center-of-mass motion of each dimer and the system becomes flexible in the sense of rigidity theory \cite{phillips1979topology, flores2010heating}, as the acoustic branch becomes a floppy mode branch due to the number of mechanical constraints being reduced by one half \cite{calladine1978, Naumis2005, kane2014topological, mao2018maxwell}. The nonlinear effects of such limit were studied for $N=3$, where the system reduces to a softened Hénon–Heiles model \cite{henon1964applicability,toledo2018escape, contopoulos2008stickiness}.    

\subsection{Nonlinear Hamiltonian written in normal modes}\label{subsec:normal_modes}

First we define the bond form factors. There are two bond types at unit cell $n$:
\begin{itemize}
  \item A bond (intra-cell) with coefficient \(\gamma_A\) and difference
    \(q_{n,2}-q_{n,1}\).
  \item B bond (inter-cell) with coefficient \(\gamma_B\) and difference
    \(q_{n+1,1}-q_{n,2}\).
\end{itemize}

Define mode-dependent difference (form-factor) for bond types, such that the bond extensions are $q_{n,2}-q_{n,1}=M^{-1/2}\sum_{k,\sigma}D_A(k,\sigma)Q_{k,\sigma}e^{ikn}$ and $q_{n+1,1}-q_{n,2}=M^{-1/2}\sum_{k,\sigma}D_B(k,\sigma)Q_{k,\sigma}e^{ikn}$:
\[
D_A(k,\sigma)=e_{k,\sigma}^{(2)}-e_{k,\sigma}^{(1)},\qquad
D_B(k,\sigma)=e^{ik}e_{k,\sigma}^{(1)}-e_{k,\sigma}^{(2)}.
\]
Since $e^{(1)}_{k,\sigma}=-\sigma s(k)/\sqrt2$ and $e^{(2)}_{k,\sigma}=1/\sqrt2$, using the explicit eigenvectors these are given by, 

\begin{equation}
D_A(k,\sigma)
=
\frac{1}{\sqrt{2}}
\left(1+\sigma s(k)\right),
\qquad
\sigma=
\begin{cases}
+1,& (+)\ \text{branch},\\
-1,& (-)\ \text{branch}.
\end{cases}
\end{equation}

and

\begin{align}
D_B(k,\sigma)
&=
-\frac{1}{\sqrt{2}}
\left(\sigma\, e^{ik}s(k)+1\right)
\nonumber\\
&=
-\frac{1}{\sqrt{2}}
\left(1+\sigma\zeta(k)\right).
\end{align}
Two useful special cases: at $k=0$ the acoustic mode (uniform translation) has $D_A=D_B=0$, while the optical mode has $D_A=\sqrt2$, $D_B=-\sqrt2$; at $k=\pi$ the acoustic mode has $D_A=0$, $D_B=-\sqrt2$ (only the weak springs are strained), while the optical mode has $D_A=\sqrt2$, $D_B=0$.
Now express the cubic potential using unit-cell count \(M=N/2\). With the momentum-conserving Kronecker delta \(\delta_{k_1+k_2+k_3,0\ (\mathrm{mod}\ 2\pi)}\), the cubic contribution ($V_3$) to the potential is given by,
\begin{equation}
\begin{split}
    V_3=\frac{1}{3\sqrt{M}}\sum_{k_1,k_2,k_3}
    \delta_{k_1+k_2+k_3,0} \times \\
    \sum_{\sigma_1,\sigma_2,\sigma_3}
    \Gamma_{\sigma_1\sigma_2\sigma_3}(k_1,k_2,k_3)\,
    Q_{k_1,\sigma_1}Q_{k_2,\sigma_2}Q_{k_3,\sigma_3}.
\end{split}
\label{eq:V3_cubic}
\end{equation}
The prefactor follows from $\sum_n e^{i(k_1+k_2+k_3)n}=M\delta$ and the factor $M^{-3/2}$ of three Bloch sums. We checked Eq.~\eqref{eq:V3_cubic} and the vertex below numerically against a direct real-space evaluation of $\frac13\sum_j\gamma_j(q_{j+1}-q_j)^3$ for random mode amplitudes, with agreement to machine precision.

The coupling coefficient is the sum of A- and B-bond contributions to
the mixing vertex:
\[
\Gamma_{\sigma_1\sigma_2\sigma_3}(k_1,k_2,k_3)
=\gamma_A \Gamma^{A} 
+\gamma_B \Gamma^{B}.
\]
where for simplicity we omit the arguments of the following functions,
\begin{equation}
\Gamma^{A}
\equiv
\prod_{j=1}^{3} D_A(k_j,\sigma_j),
\end{equation}
\begin{equation}
\Gamma^{B}
\equiv
\prod_{j=1}^{3} D_B(k_j,\sigma_j).
\end{equation}

Using $\gamma_B=\eta\gamma_A$ we find,
\begin{equation*}
|\Gamma_{\sigma_1\sigma_2\sigma_3}(k_1,k_2,k_3)|^{2}=
\gamma_A^2
\left[
|\Gamma^{A}|^2
+
2\eta\,\mathrm{Re}
\!\left(
\Gamma^{A} (\Gamma^{B})^{*}
\right)
+
\eta^2 |\Gamma^{B}|^2
\right].
\end{equation*}

The explicit formulas for $\Gamma_{\sigma_1\sigma_2\sigma_3}(k_1,k_2,k_3)$  are given in the appendix. Therein, a symmetry analysis of the vertex is made and some relevant plots and limiting cases are presented.  

Finally, collecting terms the Hamiltonian becomes:
\begin{equation}
    \begin{split}
    H &= \sum_{k,\sigma}\left(\frac{|P_{k,\sigma}|^2}{2m}+\frac{1}{2}m\omega_{k,\sigma}^2\,|Q_{k,\sigma}|^2\right) \\
    &\quad + \frac{1}{3\sqrt{M}}\sum_{\substack{k_1+k_2+k_3=0}} \sum_{\sigma_1,\sigma_2,\sigma_3}
    \Gamma_{\sigma_1\sigma_2\sigma_3}(k_1,k_2,k_3) \\
    &\qquad \times Q_{k_1,\sigma_1}Q_{k_2,\sigma_2}Q_{k_3,\sigma_3}.
    \end{split}
    \label{eq:HinNormalModes}
\end{equation}

\subsection{Observables and metrics}\label{subsec:observables}

To quantify thermalization, we monitor the energy distribution among modes and its evolution toward equipartition. Following \cite{reiss_metastable_2023}, the energy in mode $k$ at time $t$ is computed by projecting onto the linear normal mode basis:
\begin{equation}
E_{k, \sigma}(t) = \frac{1}{2} \left[ |P_{k, \sigma}|^2 + \omega_{k, \sigma}^2 |Q_{k,\sigma}|^2 \right].
\end{equation}
In the weak nonlinearity regime ($\epsilon \ll 1$, where $\epsilon = H/N$ is the energy density), the sum $\sum_{k} E_{k}(t)$ accounts for the vast majority of the total energy, permitting a consistent entropy definition. The spectral entropy $S(t)$ is defined as
\begin{equation}
S(t) = - \sum_{k, \sigma} p_{k,\sigma}(t) \log p_{k,\sigma}(t), \quad p_{k, \sigma}(t) = \frac{E_{k, \sigma}(t)}{\sum_{j, \sigma '} E_{j,\sigma '}(t)}.
\end{equation}
This quantity measures the spread of energy across modes: $S=0$ for single-mode excitation and $S=\log N$ for equipartition. In FPUT-type systems, $S(t)$ typically exhibits strong oscillations due to recurrences, obscuring long-term trends. To mitigate this, we employed a local time-averaging procedure \cite{wang_wave-turbulence_2020}:
\begin{equation}
    \bar{E}_{k, \sigma}(t) = \frac{1}{(1-\mu)t}\int_{\mu t}^{t}E_{k, \sigma}(s)\,ds,
    \label{eq:Average_energy_per_normal_mode}
\end{equation}
with $\mu = 2/3$ chosen to balance smoothing and temporal resolution \cite{benettin_time-scales_2011, benettin_christodoulidi_ponno2013}. The corresponding time-averaged spectral entropy is
\begin{equation}
  \bar{S}(t) = - \sum_{k, \sigma} \bar{e}_{k, \sigma}(t)\ln{\bar{e}_{k, \sigma}(t)},
  \label{eq:spectral_entropy_averaged_energies}
\end{equation}
where $\bar{e}_{k, \sigma}(t) = \bar{E}_{k, \sigma}(t) / \sum_{j, \sigma '} \bar{E}_{j, \sigma '}(t)$ is the time-averaged normalized mode energy [cf.~Eq.~\ref{eq:Average_energy_per_normal_mode}]. This smoothed entropy facilitated the identification of plateaus and gradual relaxation toward equipartition. Throughout, we use thermalization in its operational FPUT sense: equipartition of the time-averaged harmonic energies, for which $\bar S(t)$ attains its maximum $\log N$. A run whose $\bar S(t)$ saturates on a plateau below this value is said to display incomplete equipartition on the simulated time scale; we make no claim about its state beyond the final integration time.  $\bar S (t)$ can grow through redistribution within the acoustic branch while the optical branch remains essentially unexcited.

We additionally compute the finite-time maximum Lyapunov exponent  (ftMLE) $\lambda(t)$
via a two-trajectory protocol with initial separation $\delta_0 = 10^{-8}$ where $\mathbf{q}(t) = (q_{1}, \dots, q_{N}, p_{1}, \dots, p_{N})$ is the $2N$-dimensional phase-space trajectory and $\delta \mathbf{q}(t_{i})$ is the deviation vector between the reference and perturbed trajectories at the time $t_{i}$ \cite{contopoulos2008stickiness, skokos2010lyapunov}.
At each time step, $\delta \mathbf{q}$ is rescaled to $\delta_0$ after computing its growth rate \cite{benettin1980lyapunov}. The running average
\begin{equation}
    \lambda(t) = \frac{1}{t}\sum_{i=1}^{t/\Delta t}
    \ln\frac{\|\delta \mathbf{q}(t_i)\|}{\delta_0}
    \label{eq:ftmle}
\end{equation}
converges to the maximum Lyapunov exponent  \cite{oseledets1968}  as $t \to \infty$. For chaotic 
trajectories $\lambda(t)$ converges to a positive constant, 
while for weakly chaotic trajectories near regular phase-space 
structures it decays as $\sim t^{-\nu}$ with $0 < \nu < 1$ \cite{benettin2018fermi}.

\subsection{Numerical implementation}\label{subsec:numerics}

Numerical integration was performed using an 8th-order symplectic Kahan-Li algorithm \cite{kahan1997composition}, implemented via the \texttt{DifferentialEquations.jl} package \cite{rackauckas2017differentialequations} in Julia. Symplectic integrators were essential for this study as they preserved the Hamiltonian structure and phase-space volume \cite{hairer2006geometric}, enabling accurate long-term evolution in quasi-integrable systems where energy conservation was critical.

A fixed timestep $\Delta t = 0.05$ was used for the $N=64,128,256$ reference runs and $\Delta t = 0.2$ for the robustness runs; being symplectic, the integrator keeps the energy error bounded without secular drift ($|\Delta E/E|\lesssim 2\times10^{-8}$ for $\Delta t = 0.2$, $N=256$) \cite{hairer2006geometric}.  
Time is reported in dimensionless units scaled by the linear period of the initially excited mode. This normalization ensures that one time unit corresponds to one oscillation cycle of the initial excitation, facilitating physical interpretation.

Two boundary conditions were examined: fixed endpoints (FBC, $q_0 = q_{N+1} = 0$) and periodic boundary conditions (PBC, $q_{N+1} = q_1$). The lowest acoustic mode is initially excited (mode 1 for FBC; mode 2 for PBC, since mode 1 is the uniform translation) and for FBC, additional runs were made for an initial condition of kicking two contiguous masses on a central dimer. The main scan covers from weak to strong dimerization. To assess finite-size effects, we ran  FBC and PBC simulations for $N=64, 128$ and $256$, with $\Delta \kappa$ below and above $\Delta \kappa = 0.5$  at fixed $\epsilon$. For the energy-density effects we ran PBC simulations sweeping values of $\alpha= 0.05, 0.1, 0.2$ at fixed energy-density $\epsilon$ which is equivalent to varying the energy density through the invariant $\bar\lambda^{2}\equiv\alpha^{2}\epsilon$ (see Appendix~\ref{app:rescaling}).  All simulations and analysis code were implemented in Julia and provided in the Supplementary Material and in a public repository in GitHub for reproducibility. 

\section{Results}\label{sec:results}

We present the results of the numerical simulations described in Sec.~\ref{subsec:numerics}. 

\subsection{Modal energy evolution for varying \texorpdfstring{$\Delta\kappa$}{Delta-kappa}}\label{subsec:heatmap}

\begin{figure*}[t]
    \centering
    \includegraphics[width=\textwidth]{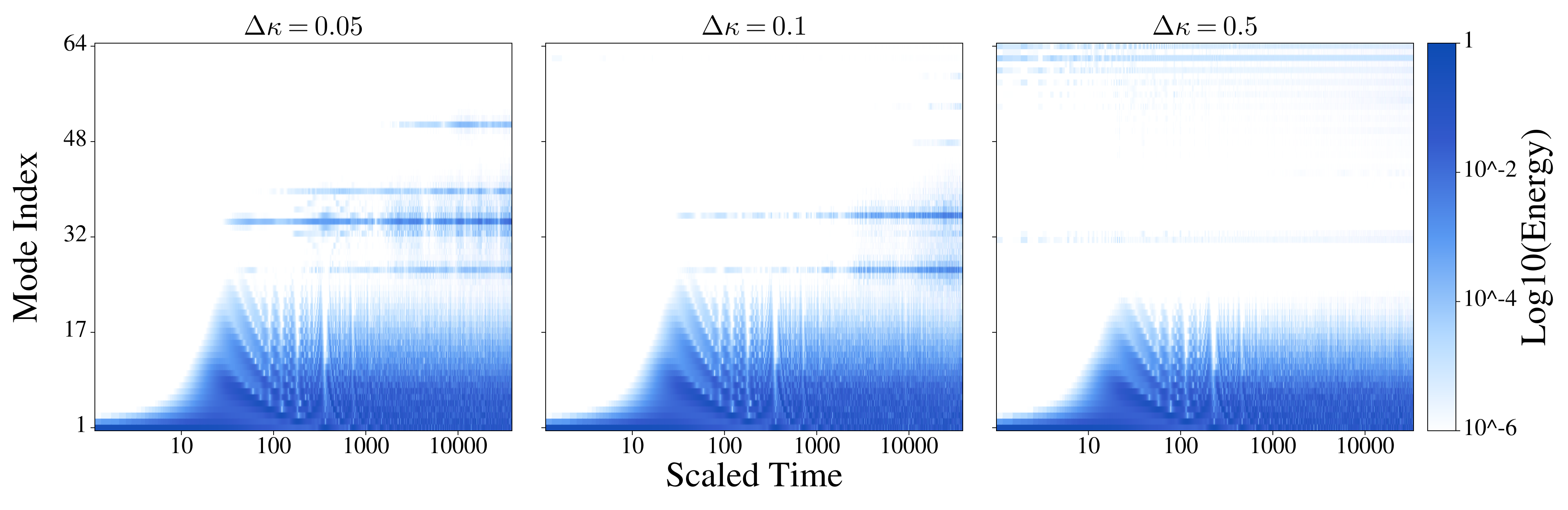}
    \vspace{0.5em}
    \includegraphics[width=\textwidth]{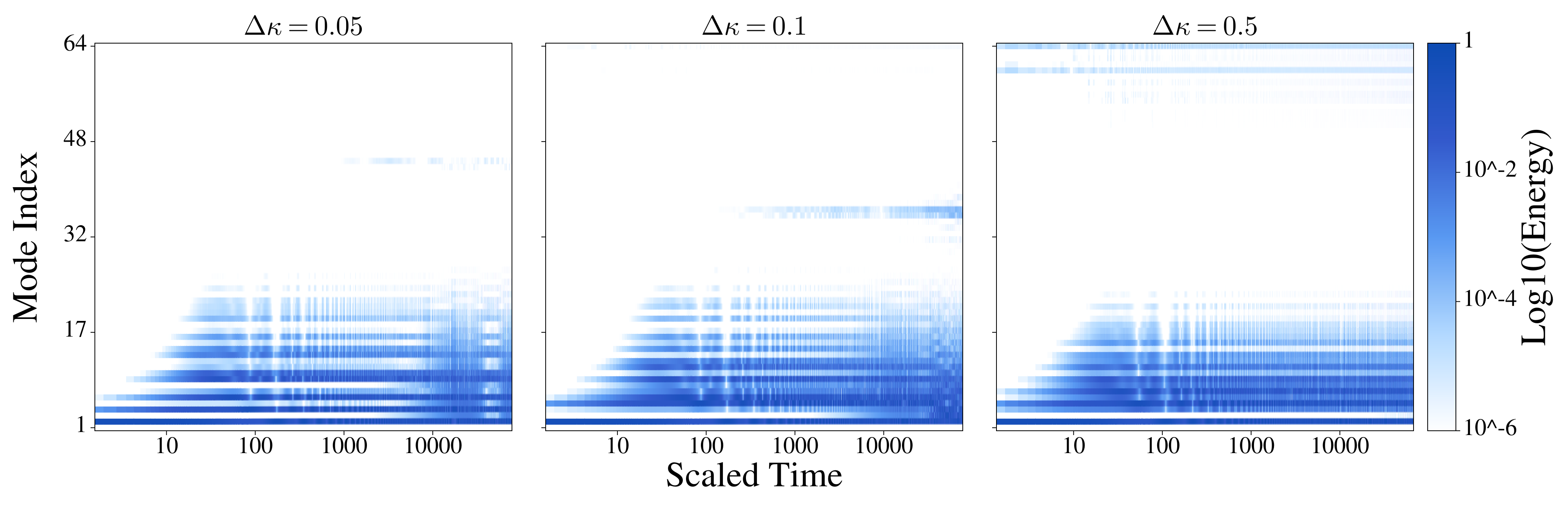}
    \caption{
  Modal energy heat maps for $\alpha$-type nonlinearity, $N=64$, $\epsilon=0.0069$, $\alpha = 0.1$.
  Top panel: FBC (first-mode excitation).
  Bottom panel: PBC (second-mode excitation). Observe how for the $\Delta \kappa=0.5$ case, the highest frequency optical mode is excited due to the first available umklapp process.}    \label{fig:HeatMap_Different_Springs_Alpha_Log_Scale}
\end{figure*}

Figure~\ref{fig:HeatMap_Different_Springs_Alpha_Log_Scale} displays the fraction of total energy per normal mode as a function of time for $\alpha$-type nonlinearity under FBC (top) and PBC (bottom). The dimerization strength $\Delta\kappa$ acts as the primary control parameter for the energy cascade pattern.

For small spring contrast ($\Delta \kappa < 0.1$), energy diffused rapidly from the initially excited low-frequency acoustic mode toward the high-frequency edge of the acoustic branch.  
As $\Delta\kappa$ increases from $0.1$ to $1$, the wider phononic gap delays the energy transfer into the optical branch. Notice in Fig.~\ref{fig:HeatMap_Different_Springs_Alpha_Log_Scale} that for $\Delta \kappa=0.5$, the highest frequency mode is first excited. As we discuss below, this is a consequence of the last available umklapp process. Also, even at the largest nonlinearity strength studied, optical modes were only marginally excited within the window of time of Fig.~\ref{fig:HeatMap_Different_Springs_Alpha_Log_Scale}. 
 
\subsection{ Spectral entropy and approach to equipartition}\label{subsec:entropy}

\begin{figure*}[!ht]
    \centering
    \begin{subfigure}{0.48\textwidth}
        \centering
        \includegraphics[width=\textwidth]{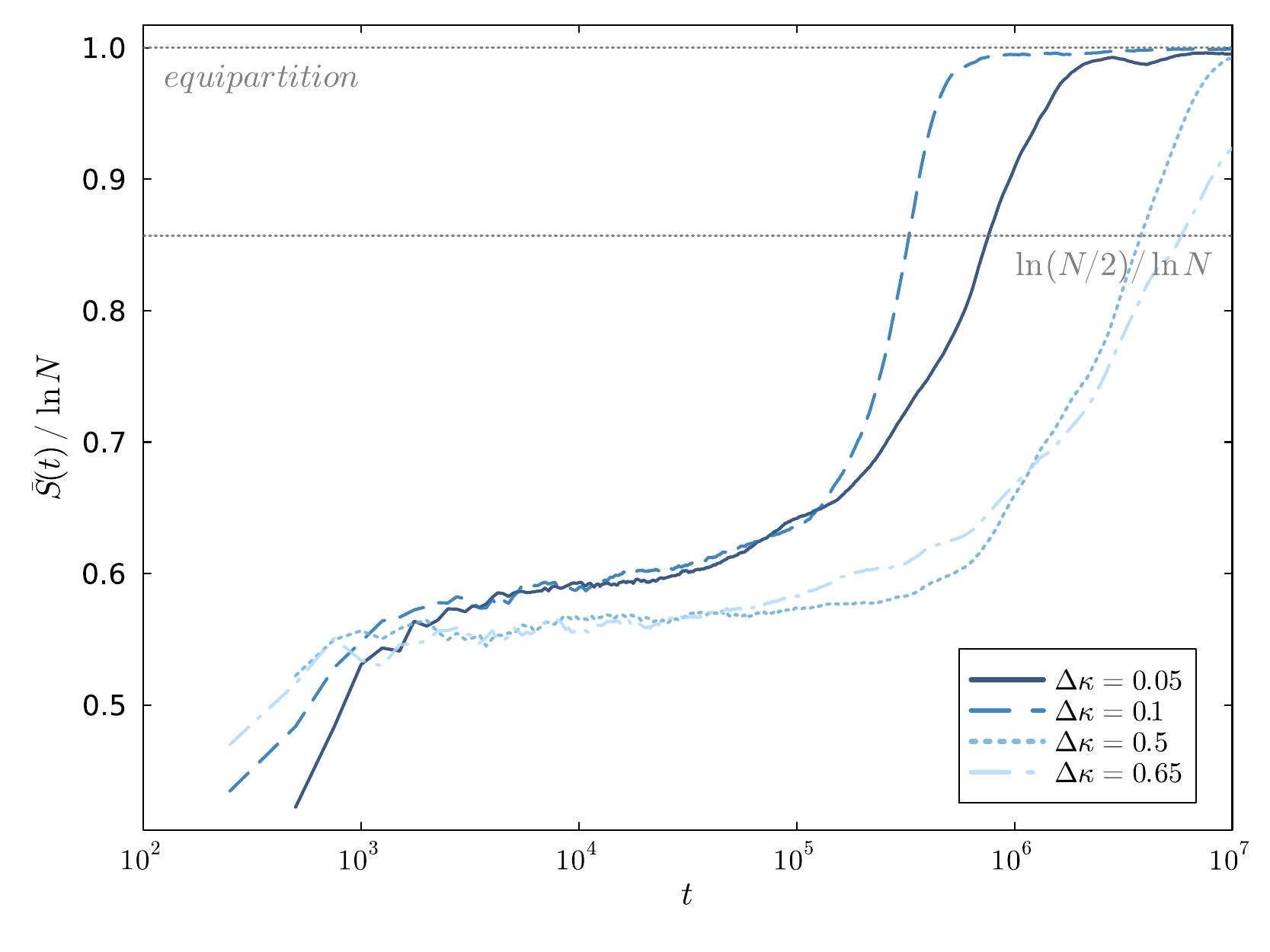}
        \caption{$N = 128$}
        \label{fig:entropy_fbc_N128}
    \end{subfigure}
    \hfill
    \begin{subfigure}{0.48\textwidth}
        \centering
        \includegraphics[width=\textwidth]{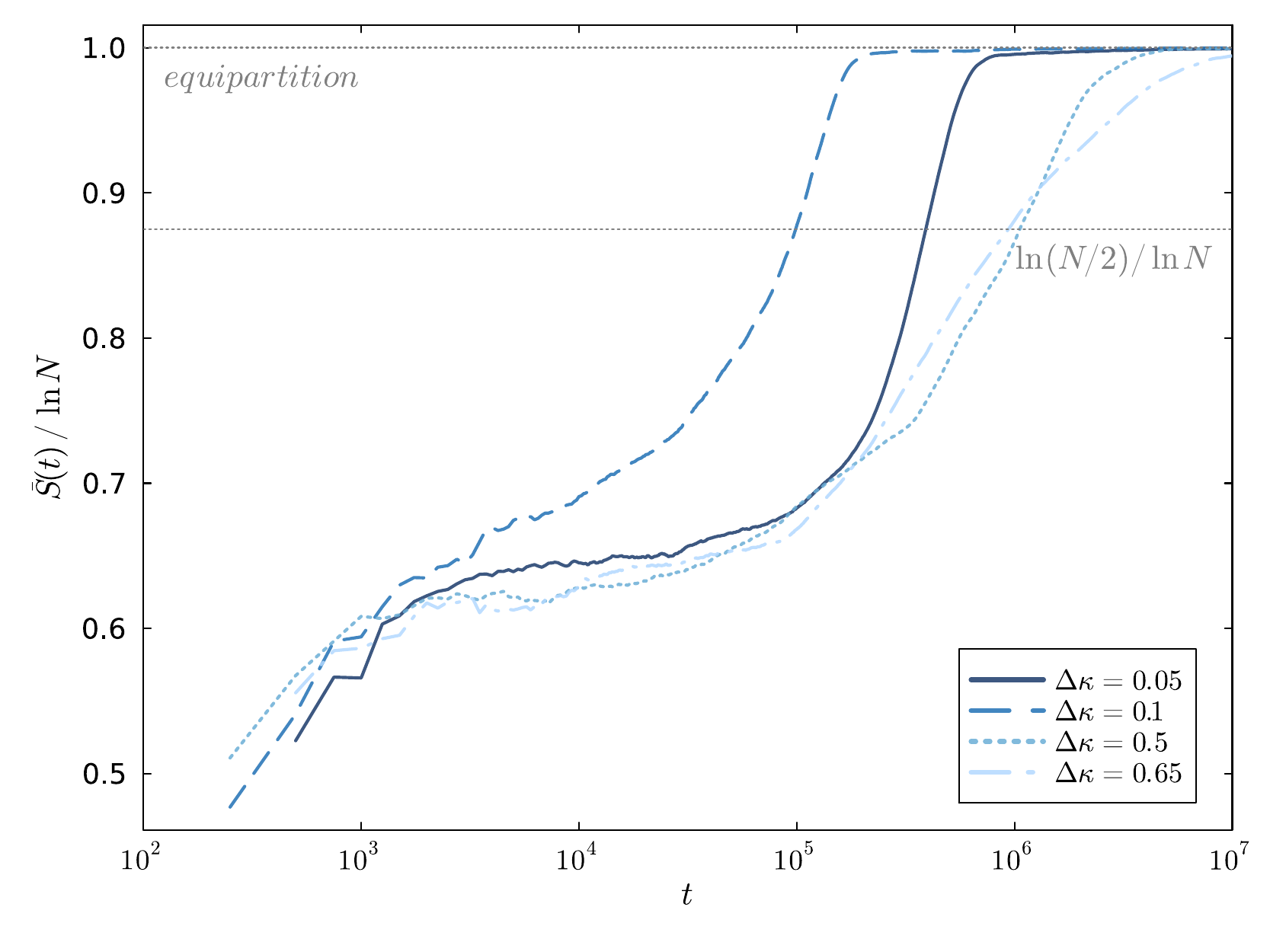}
        \caption{$N = 256$}
        \label{fig:entropy_fbc_N256}
    \end{subfigure}
    \caption{Time-averaged spectral entropy $\bar{S}(t)$ divided by the value of $\ln{N}$ (which indicates equipartition) vs.\ time for FBC for different values of $\Delta \kappa$ using as initial condition a first-mode excitation. The system sizes are (a) $N = 128$ and (b) $N = 256$. For $\Delta \kappa = 0.05$ and $0.1$ the entropy leaves the initial plateau around $t \sim 10^{5}$ and reaches equipartition around $t\sim 10^6$ and earlier for $N=256$. At and above the threshold ($\Delta\kappa = 0.5, 0.65$) the rise is delayed compared to previous cases. Dotted horizontal lines indicate the entropy values at equipartition and when half of the modes are excited ($\ln(N/2)/\ln N$). }
    \label{fig:entropy_fbc_size}
\end{figure*}

The time-averaged spectral entropy $\bar{S}(t)$ [Eq.~\ref{eq:spectral_entropy_averaged_energies}] provides a single-number summary of how broadly the energy is distributed across the $N$ normal modes. Figure~\ref{fig:entropy_fbc_size} shows $\bar{S}(t)/\ln N$ for two sizes of the same system, $N=128$ and $N=256$, under fixed boundary conditions, for $\Delta\kappa \in \{0.05,\,0.1,\,0.5, 0.65\}$ and with the same initial condition: the first normal mode is excited. Two horizontal lines are shown in both panels: the first one indicates the entropy reached when the energy is shared equally among half of the normal modes, $\bar{S}/\ln N = \ln(N/2)/\ln N$, and the second indicates full equipartition, where $\bar{S}/\ln N$ reaches its maximum value of $1$. In both sizes, $\bar{S}(t)$ first shows a relaxation stage that ends in a plateau at $\bar{S}/\ln N \approx 0.6$ for $t \sim 10^{3}$. In all cases, the system eventually approaches equipartition within $t= 10^ {7}$, except $\Delta \kappa = 0.65$ at $N=128$, but showing an approach to equipartition. The delay shown above the threshold of $\Delta \kappa$ decreases with system size.  

\begin{figure*}[!ht]
    \centering
    \begin{subfigure}{0.48\textwidth}
        \centering
        \includegraphics[width=\textwidth]{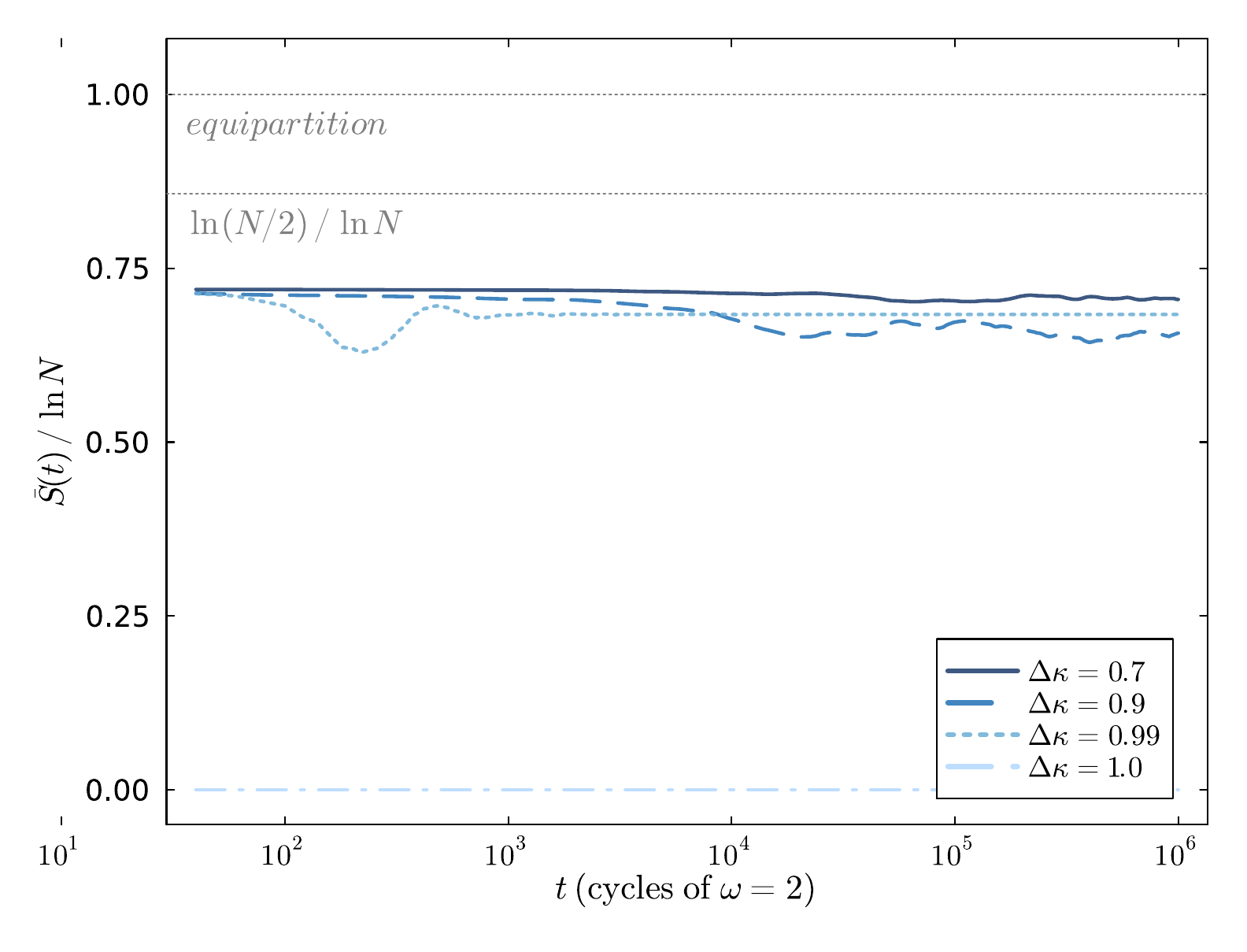}
        \caption{$N = 128$}
        \label{fig:dimer_fbc_N128}
    \end{subfigure}
    \hfill
    \begin{subfigure}{0.48\textwidth}
        \centering
        \includegraphics[width=\textwidth]{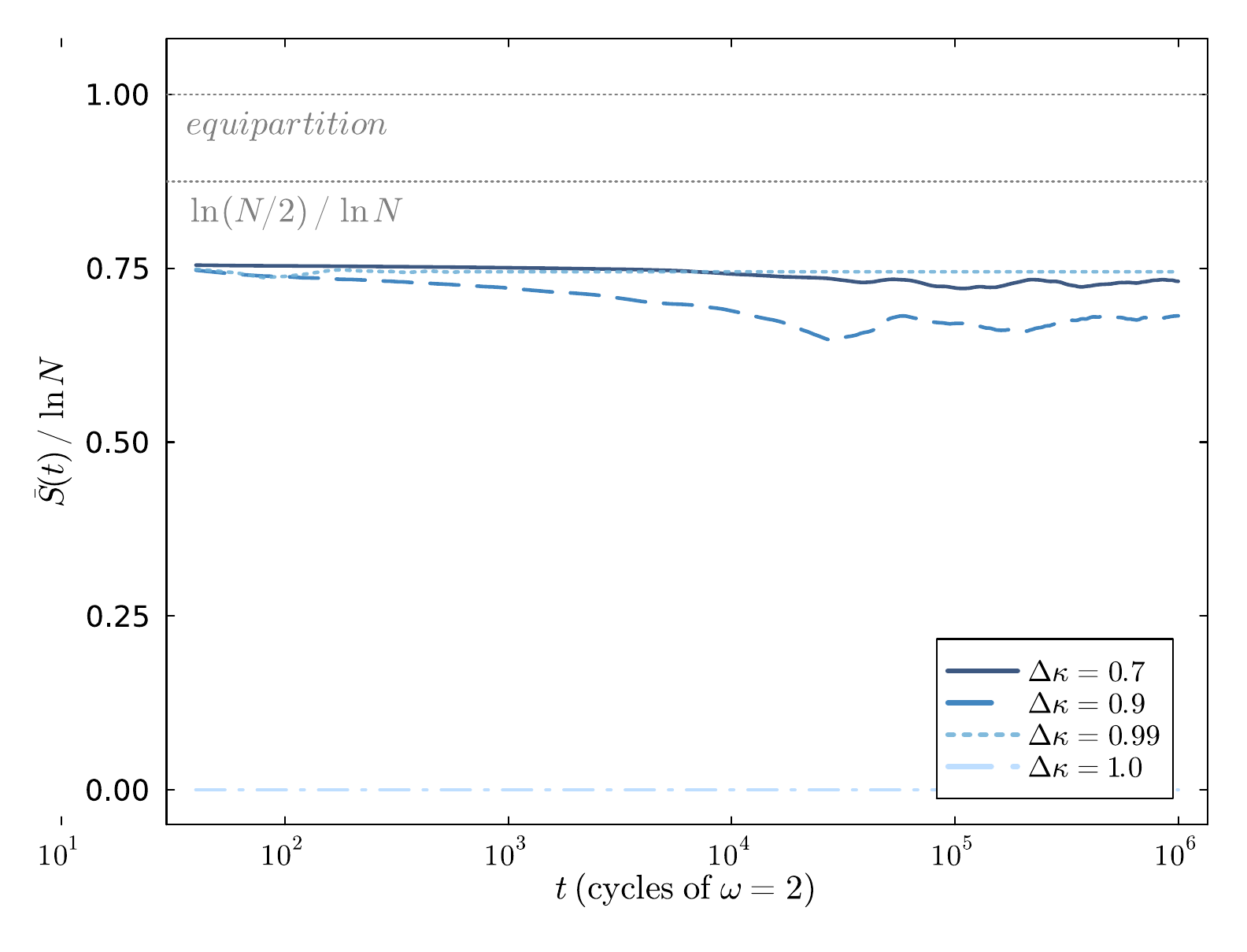}
        \caption{$N = 256$}
        \label{fig:dimer_fbc_N256}
    \end{subfigure}
    \caption{$\bar{S}(t) / \ln{N}$ vs.\ time for FBC using
    a dimer excited at the center of the chain for $\alpha = 0.1$, $\epsilon = 0.0069$ and two different sizes, (a) $N = 128$ and (b) $N = 256$. The central dimer is excited using opposite velocities for each mass, resulting in zero initial displacement and zero center-of-mass momentum. Only values of $\Delta \kappa $ above the three-wave threshold 
    $\Delta\kappa > 0.5$ are shown.}
    \label{fig:dimer_fbc_size}
\end{figure*}

Initial and boundary conditions also play a fundamental role in whether the system eventually reaches equipartition or remains trapped without completely thermalizing.  For example, in the limit $\Delta \kappa \rightarrow 1$, the normal modes approach degeneracy and thus almost any linear combination of them are close to a normal mode solution. For $\Delta \kappa =1$, the normal modes are either localized on dimers or linear combinations of them. Therefore, in this regime it is interesting to excite only a dimer at the center of the chain. Fig. \ref{fig:dimer_fbc_size} presents $\bar{S}(t)$ for two different sizes of the same system with different values of $\Delta \kappa$ in the interval $(1/2,1]$ using as initial condition an excitation of a dimer in the center of the chain. Fig. \ref{fig:dimer_fbc_size} shows that the equipartition is never achieved, and for the case $\Delta \kappa=1$, as expected, the dimer mode never relaxes as it is disconnected from the chain. Clearly phonon diffusion is very much affected in this regime.

\begin{figure}[ht]
    \centering
    \includegraphics[width=\linewidth]{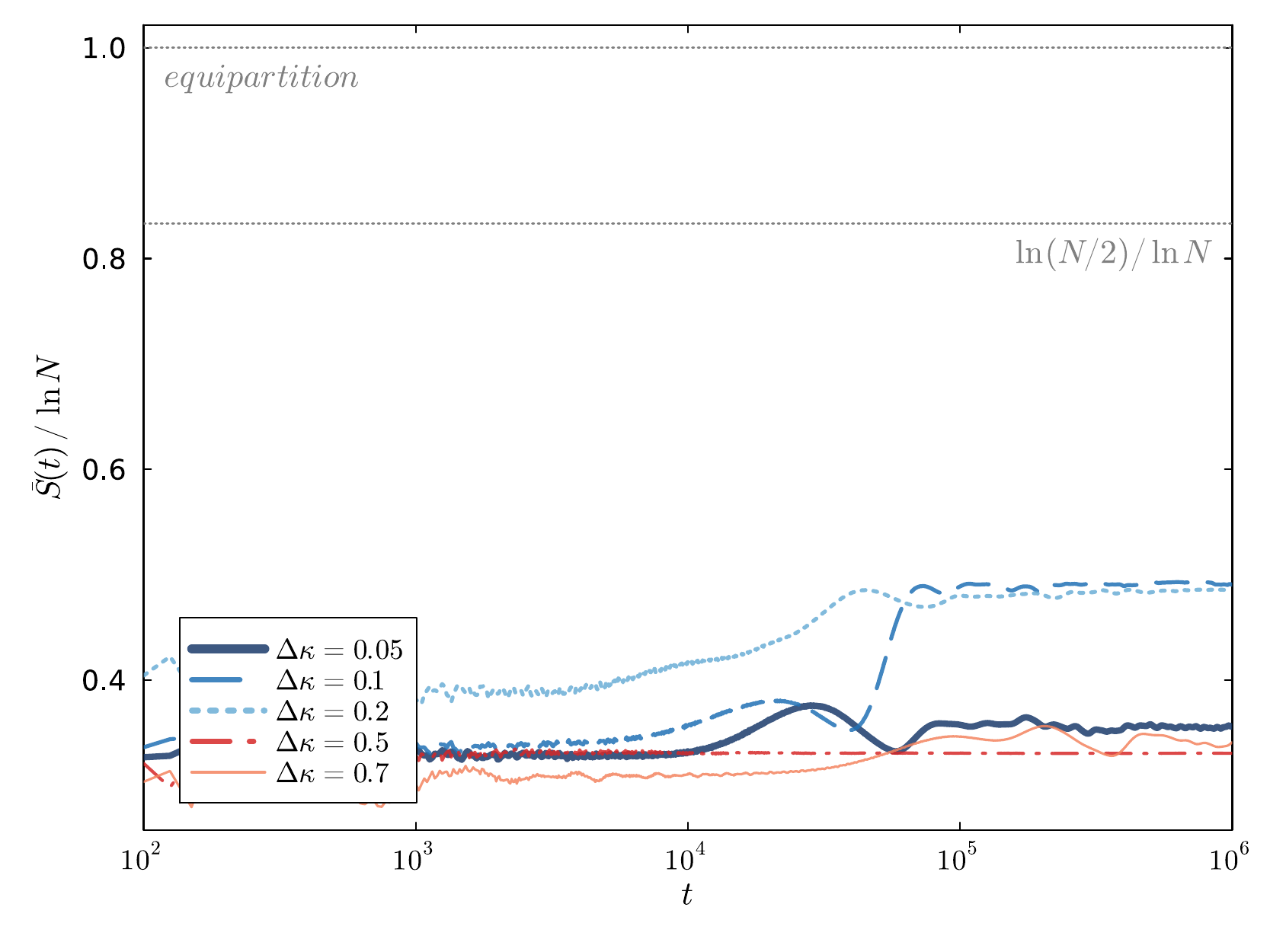}
     \caption{Time-averaged spectral entropy $\bar{S}(t)/\ln{N}$ vs.\ time for PBC,
    second-mode excitation, $\alpha = 0.1$, $N = 64$, $\epsilon = 0.0069$.
    Among sub-threshold values (thick blue), $\Delta\kappa = 0.2$ shows the largest entropy growth, followed by $\Delta\kappa = 0.1$ and $\Delta\kappa = 0.05$,
    suggesting that the effective transference of energy
    depends non-trivially on dimerization within the open-channel regime.
    Above the threshold ($\Delta\kappa = 0.5$ and $0.7$, thin red),
    the system remains mostly trapped in low-entropy plateaus throughout
    the simulation window ($t=10^6$).  }
    \label{fig:SpectralEntropy_Different_Springs_Periodic_Conditions_N64_alpha}
\end{figure}

\begin{figure}[t]
    \centering
    \includegraphics[width=\linewidth]{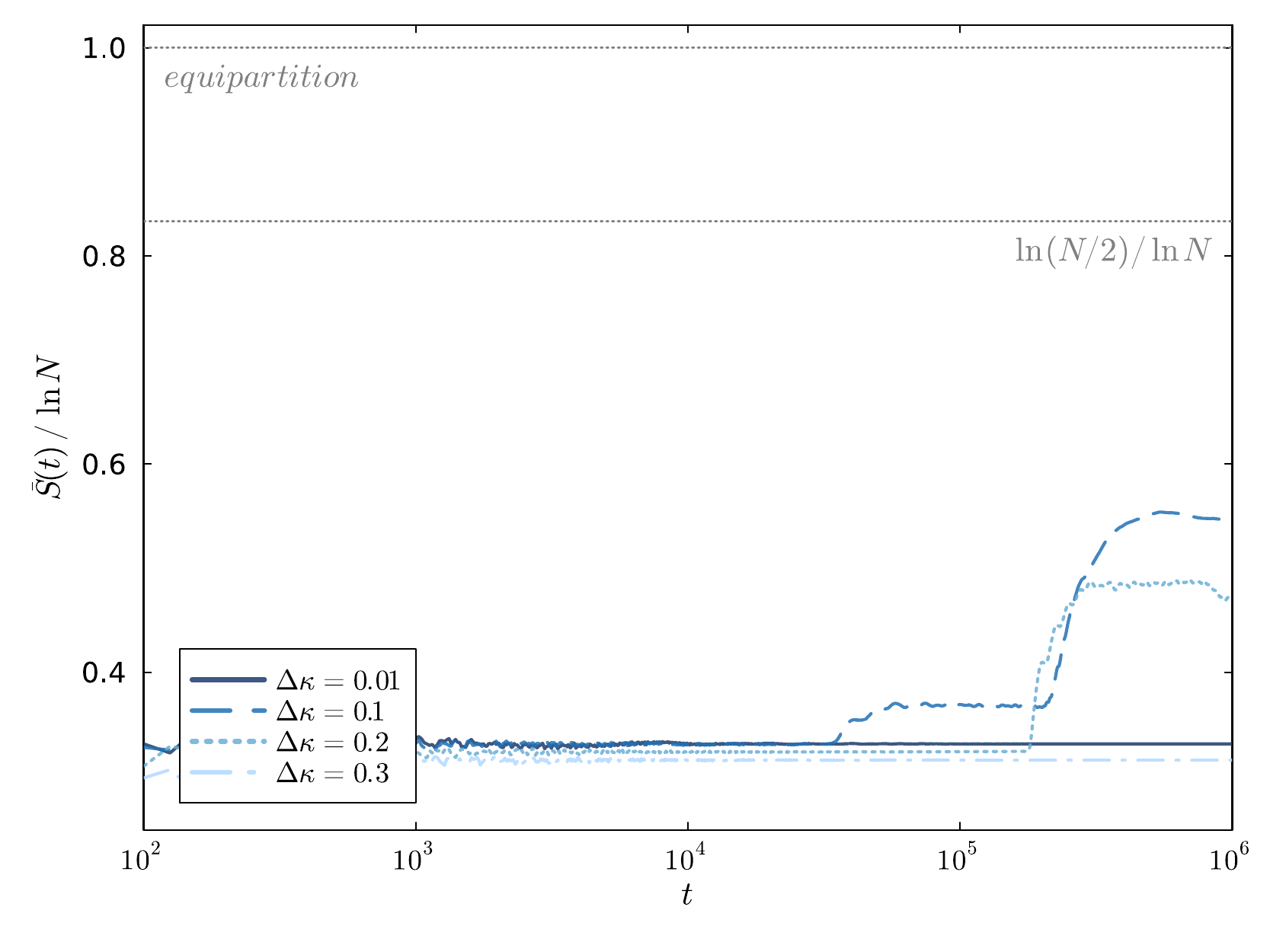}
    \caption{$\bar{S}(t)/\ln N$ for FBC with second-mode excitation, $\alpha = 0.1$, $N=64$, $\epsilon = 0.0069$. Entropy grows only for $\Delta\kappa=0.1$ and $0.2$, with late jumps at $t\approx4\times10^{4}$ and $2\times10^{5}$; $\Delta\kappa=0.01$ and $0.3$ stay on plateaus near $0.3$ up to $t=10^{6}$. All values remain well below those reached with first-mode excitation (Fig.~\ref{fig:entropy_fbc_size}). } 
     \label{fig:SpectralEntropy_Different_Springs_Fixed_Conditions_N64_alpha_second_mode}
\end{figure}

Fig.\ref{fig:SpectralEntropy_Different_Springs_Periodic_Conditions_N64_alpha} shows the system at PBC at different values of $\Delta \kappa$. For values below the threshold $\Delta \kappa = 0.05, 0.1$ different behaviors of the entropy are present. In the regime near of the monoatomic case $\Delta \kappa = 0.05$, the systems presents relaxation without showing signs of a rise in entropy at the shown periods of time but for $\Delta \kappa = 0.1, 0.2$ we can see a rise in entropy around $t= 10^4 - 10^5$. In the threshold of $\Delta \kappa = 0.5$ we found that the system reaches relaxation without fully thermalizing at $t=10^6$.\ref{fig:SpectralEntropy_Different_Springs_Fixed_Conditions_N64_alpha_second_mode} shows the first case of FBC but exciting the second normal mode of the system as the initial condition. We can see that, contrasting with Fig. \ref{fig:entropy_fbc_size}, the reached values of entropy are lower than the first-mode excited case, although the entropy shows plateaus. As in the PBC case, near the monoatomic limit ($\Delta\kappa = 0.01$) the entropy remains on a plateau, whereas for $\Delta\kappa = 0.1$ and $0.2$ it rises, without reaching equipartition within the simulated time. All of these results show that the thermalization is a complex process depending on different conditions. Nevertheless, we can find a common behavior; In different variations of the system we can see an earlier increase of spectral entropy around values of $\Delta \kappa$ below $0.5$ and how the systems seems to be trapped for longer times for values above the threshold of $\Delta \kappa = 0.5$, giving us an idea of how the presence of a gap in the system can control the thermalization of the systems, delaying or accelerating interbranch energy transfer.

\subsection{\texorpdfstring{Energy-density scaling}{Energy-density scaling}}\label{subsec:scaling}
\begin{figure*}[t]
  \centering
  \includegraphics[width=\textwidth]{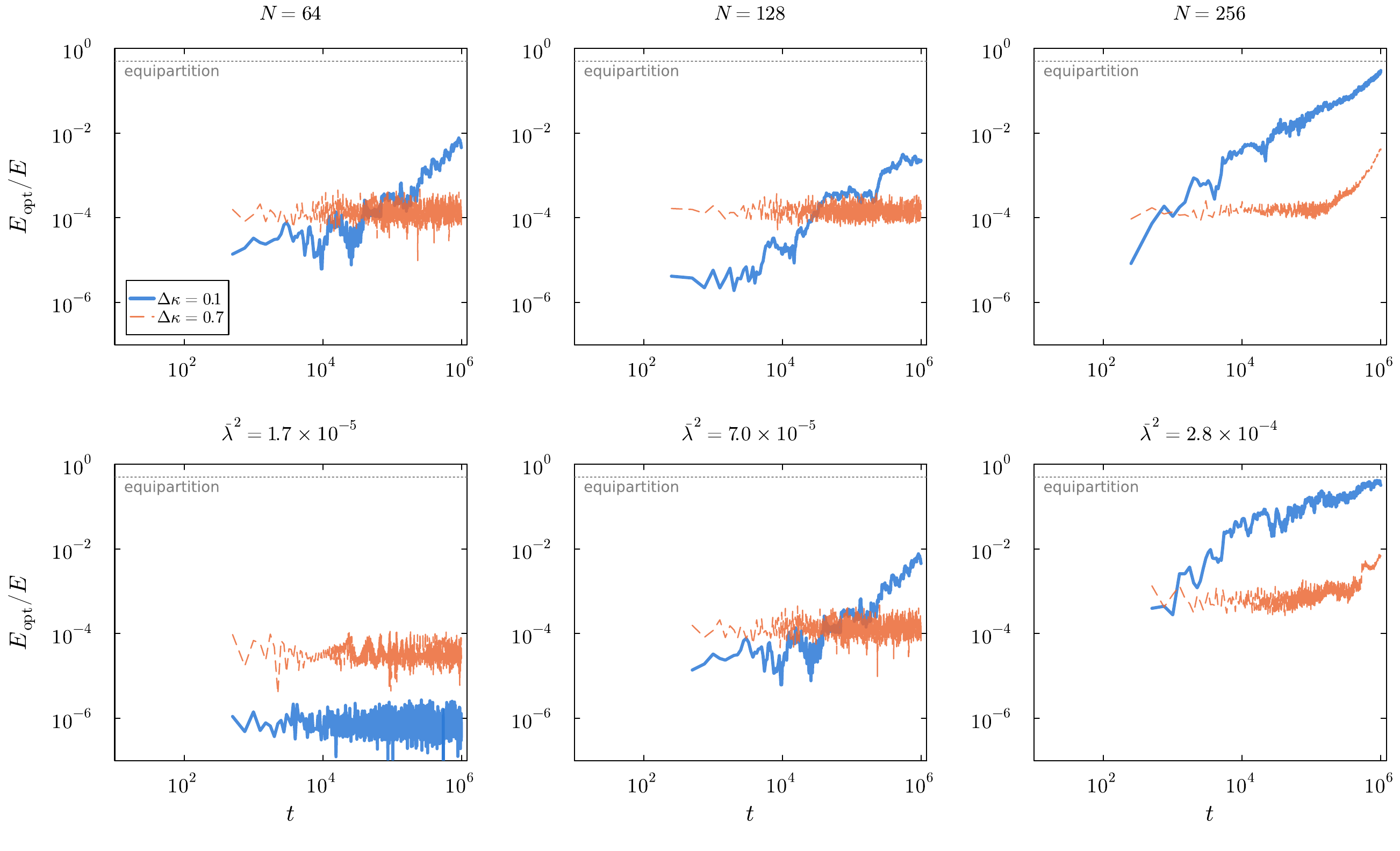}
  \caption{Optical energy fraction $E_{\mathrm{opt}}/E$ vs.\ time below ($\Delta\kappa=0.1$, solid) and above ($\Delta\kappa=0.7$, dashed) the three-wave channel threshold, with PBC and second-mode excitation. Top: $N=64,128,256$ at $\epsilon=0.0069$, $\alpha=0.1$. Bottom: $N=64$ at $\bar\lambda^{2}=\alpha^{2}\epsilon$ for $\alpha=0.05,0.1,0.2$ (equivalent to varying $\epsilon$, see Appendix~\ref{app:rescaling}). The dotted line marks equipartition.}
  \label{fig:interband_transfer}
\end{figure*}

Fig. \ref{fig:interband_transfer} shows the fraction of total energy transferred to the energy of the optical modes, giving us a way to measure interband energy transference for different sizes of the system and at different values of energy density $\epsilon$ using the scaling parameter $\bar{\lambda}^{2} = \alpha^2 \epsilon$ (see Appendix \ref{app:rescaling}). The lower panel in particular shows an increase of the fraction of energy in optical modes as the parameter $\bar{\lambda}^2$ increases and compared with two values of $\Delta \kappa$, below and above the derived threshold of $\Delta \kappa = 0.5$. Specifically, we notice that at increasing values of $\bar{\lambda}^{2}$ interbranch energy transfer is faster for the value of $\Delta \kappa$ below a threshold, in this case $\Delta \kappa = 0.1$. In other words, increasing the density of energy and fixing $\Delta \kappa$ for a small value below the threshold $\Delta \kappa = 0.5$  enhances interbranch energy transfer, helping lastly to reach equipartition faster.

\section{Discussion}\label{sec:discussion}

Two main ingredients govern the thermalization process. The first is the existence of a resonant manifold, and the second is the magnitude of the interaction vertex associated with the resonance. We first discuss the most important resonant manifold and then examine the vertex behavior.

\subsection{Three-wave resonance conditions and threshold}\label{subsec:resonances}

\begin{figure}
    \centering
    \includegraphics[width=1.0\linewidth]{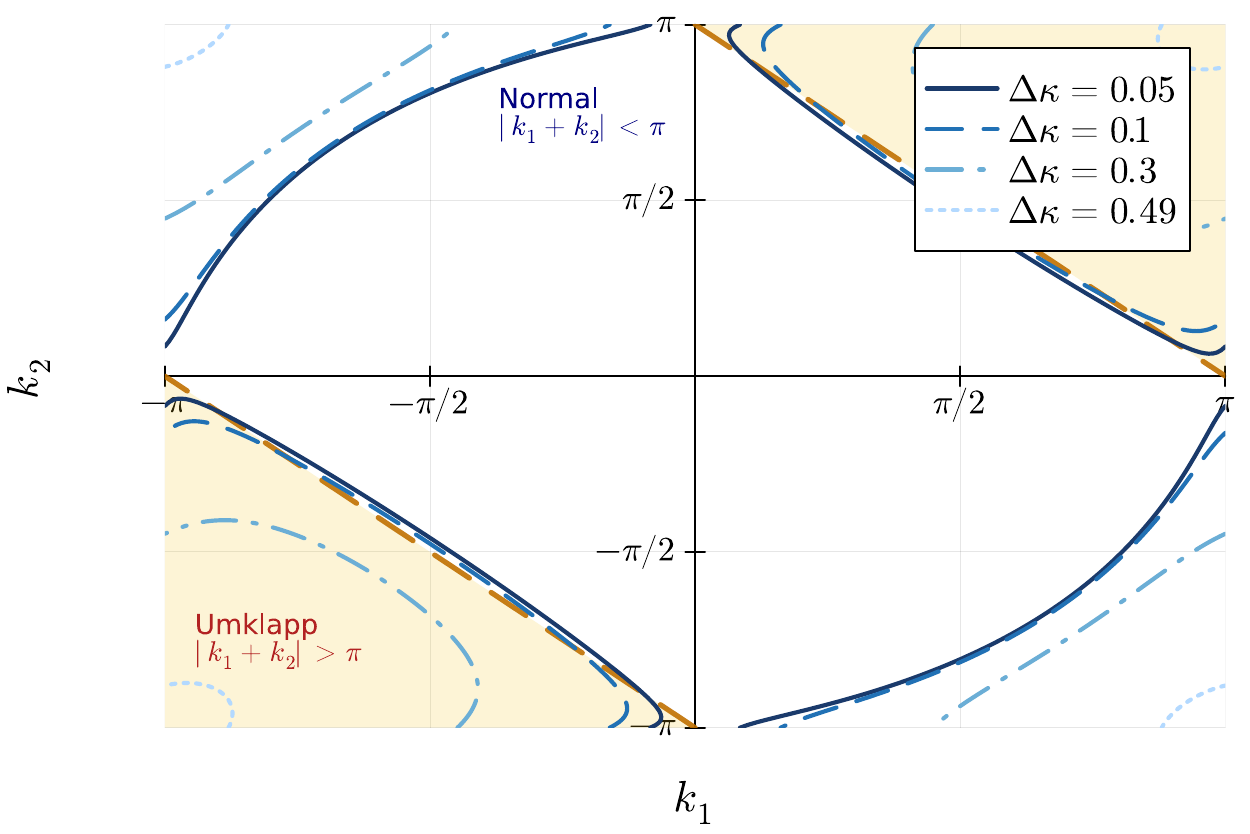}
    \caption{Contour plot of the resonance condition manifold $|\omega_-(k_1)+\omega_-(k_2)-\omega_+(k_3)|<\Lambda$ for  $\Lambda\ll1$ for different  $\Delta\kappa$ values and for the $\alpha$-type nonlinearity. This shows how the accessible resonant triads depend on $\Delta\kappa$. The momentum conservation and gap width controls the density of resonances connecting acoustic and optical modes. The resonance manifold shrinks onto the zone corner $k_1=k_2=\pi$ and disappears as $\Delta \kappa \to 0.5$, no three-wave resonance exists for $\Delta\kappa > 0.5$. The value $\Delta \kappa= 0.5$ signals the  umklapp process ($k_1=k_2=\pi\to k_3=0$) to be allowed. The shaded area corresponds to the possible umklapp processes allowed by momentum conservation ($|k_1+k_2|>\pi$). The red dotted lines indicate the limits of such region.}
    \label{fig:resonance_level_set_alpha}
\end{figure}

Equation~\eqref{eq:HinNormalModes} shows that the nonlinearity couples the normal modes, and whether a given triad of modes exchanges energy secularly is decided by two conditions. The first is exact: the Kronecker delta in the cubic potential $V_3$ [Eq.~\eqref{eq:V3_cubic}] imposes
\begin{equation}
    k_{1} + k_{2} + k_{3} = G,
    \label{eq:conservation_of_momentum_condition}
\end{equation}
with $G=0,\pm 2\pi, \dots$ a reciprocal-lattice vector (momenta are defined per unit cell). It follows directly from the translational symmetry of the lattice and holds at any amplitude. 
As seen in Fig.~\ref{fig:resonance_level_set_alpha}, for $\Delta \kappa \rightarrow 0$, i.e., $\eta \rightarrow 1$, the manifold is close to the lines $k_1+k_2=\pm \pi$. These lines define the border between umklapp and normal processes. In Fig. \ref{fig:resonance_level_set_alpha} we shade the umklapp area ($|k_1+k_2|>\pi$). At $\Delta\kappa=0$ the resonant set does not disappear kinematically: it is the folded image of the monoatomic dispersion. What vanishes is the coupling.
Notice that the momentum condition Eq. (\ref{eq:conservation_of_momentum_condition}) distinguishes normal processes ($G=0$) from Umklapp processes ($G\neq 0$), in which the sum is remapped onto the first Brillouin zone by a reciprocal-lattice vector.

 The second is a result of weak-nonlinearity theory~\cite{l2010discrete, nazarenko_wave_2011, lvov_double_2018}: in the small-amplitude regime ($\epsilon \ll 1$), non-resonant terms oscillate rapidly and average to zero over long times, leaving only processes that are resonant \cite{nazarenko_wave_2011,bustamante2019exact,wang2024thermalization, lin_fu_wang_zhang_zhao_2025},
\begin{equation}
    k_1+...+k_L=k_{L+1}+..+k_{L+T}  
\end{equation}
\begin{equation}
 \omega_{\sigma_1}(k_1)+...+\omega_{\sigma_L}(k_L)=\omega_{\sigma_{L+1}}(k_1)+...+\omega_{\sigma_{L+T}}(k_L)
\end{equation}
which convert $S$ waves into $T$ waves as the drivers of secular energy redistribution among modes. The order of the resonance is given by $S+T$. In this work, we will concentrate on the lowest order resonance, 
\begin{equation}
    \omega_{\sigma_1}(k_1) \pm  \omega_{\sigma_2}(k_2) \pm  \omega_{\sigma_3}(k_3)=0
    \label{eq:freq_resonance}
\end{equation}
Other high order resonances are present producing long time relaxation, yet in this work we separate the most important contribution: the three wave channel. For small size systems, the exact three wave resonances are replaced by quasiresonances:
\begin{equation}
    |  \omega_{\sigma_1}(k_1) \pm  \omega_{\sigma_2}(k_2) \pm  \omega_{\sigma_3}(k_3)|\le \Lambda,
    \label{eq:quasiresonance_condition}
\end{equation} 
where $\Lambda$ is a resonance width determined by the system's finite size. Here the structure of the dispersion relation changes the spectral constraints on which branch combinations $(\sigma_1,\sigma_2,\sigma_3)$ can satisfy Eq.~\eqref{eq:freq_resonance}.
As detailed in the appendix, combined with the conservation of energy and momentum conditions, we find that, of the 8 combinations of $(\sigma_{1}, \sigma_{2}, \sigma_{3})$ (read as the fusion $1+2\to3$), four are impossible by the band bounds alone. The channel $(-,-,-)$ is in principle impossible by the subadditivity of the acoustic branch. However, as  $\Delta \kappa \rightarrow 1$ the frequency dispersion becomes flat and the modes become quasiresonant, helping for a fast population of the modes. $(-,-,+)$ is the only open channel that converts two acoustic phonons into an optical one, and $(-,+,+)$, $(+,-,+)$ (equivalent under exchange of the two incoming phonons) require an optical phonon in the initial state. The most important contribution for energy transfer of the optical branch is the acoustic-acoustic to optical branch ($\sigma_1=-,\sigma_2=-,\sigma_3=+$). The proofs are given in Appendix~\ref{app:channels}. Fig. \ref{fig:resonance_level_set_alpha} shows the manifold corresponding to the most important channel for optical mode excitation ($\sigma_1=-,\sigma_2=-,\sigma_3=+$). The curves in Fig. \ref{fig:resonance_level_set_alpha} give  the geometric set of momentums $k_1,k_2$ which satisfy

 \begin{equation}
    |  \omega_{-}(k_1) +  \omega_{-}(k_2) - \omega_{+}(-k_1-k_2)|\le \Lambda,
\end{equation} 
for a given very small $\Lambda$ value at different $\Delta \kappa$ values. This illustrates how the density of resonant triads connecting acoustic and optical modes decreases sharply as $\Delta\kappa$ increases and reinforces the interpretation that the phononic gap acts as a tunable filter for nonlinear interactions controlling the thermalization pathway. The figure shows that for $\Delta \kappa \leq 0.5$ there are many resonant triads connecting the acoustic and optical branches, while for $\Delta \kappa > 0.5$ the gap is too wide to allow any resonant triads, suppressing three-wave transfer between the two sectors. Momentum conservation also plays a role. Both conclusions arise from Eq.~\eqref{eq:dispersion_relation_different_springs_bonita}, as the acoustic and optical bands are bounded by
\begin{align}
    0 &\leq \omega_-(k) \leq \omega_-^{\max} = \sqrt{2(1-\Delta\kappa)},
    \label{eq:ac_bounds}\\
    \omega_+^{\min} &= \sqrt{2(1+\Delta\kappa)} \leq \omega_+(k) \leq 2,
    \label{eq:opt_bounds}
\end{align}
The band bounds alone only give the necessary condition $2\omega_-^{\max}\ge\omega_+^{\min}$, i.e.\ $\Delta\kappa\le3/5$; the sharper threshold comes from momentum conservation. A necessary condition for an acoustic-acoustic-optical resonance is that two acoustic phonons can sum to an optical frequency. The extremal case occurs at $k_{1} = k_{2} = \pi$ where the acoustic branch reaches its maximum, $\omega_{-}^{max} = \sqrt{2(1-\Delta \kappa
)}$ and momentum conservation fixes the third wavevector to $k_{3} = 0 \hspace{1mm}( \text{mod}  \hspace{1mm}2\pi )$, so the relevant optical frequency is $\omega_{+}(0) = \omega_{+}^{max} = 2$.
Imposing 
$$ \omega_{+}(0) = \omega_+^{max} \leq 2\omega_-^{max} = 2\omega_{-}(\pi) $$ 
gives 
$$\Delta\kappa \leq \frac{1}{2}.$$
That the corner $(k_1,k_2)=(\pi,\pi)$ maximizes,
\begin{equation}
 F(k_1,k_2)=\omega_-(k_1)+\omega_-(k_2)-\omega_+(k_1+k_2)   
\end{equation}
over the whole zone (so that $F\ge0$ somewhere if and only if $F(\pi,\pi)\ge0$) was verified numerically for $\Delta\kappa\in[0.3,0.6]$ on a $1200^2$ grid: $\max F=2\sqrt{2(1-\Delta\kappa)}-2$, vanishing at $\Delta\kappa=1/2$ and negative beyond.
This threshold coincides with the first allowed umklapp process: the merging of two zone-boundary acoustic phonons $(k = \pm \pi$ zero group velocity) into a zone-center optical phonon ($K=k_1+k_2+k_3=2\pi$).

\begin{figure}[!ht]
    \centering
    \includegraphics[width=\linewidth]{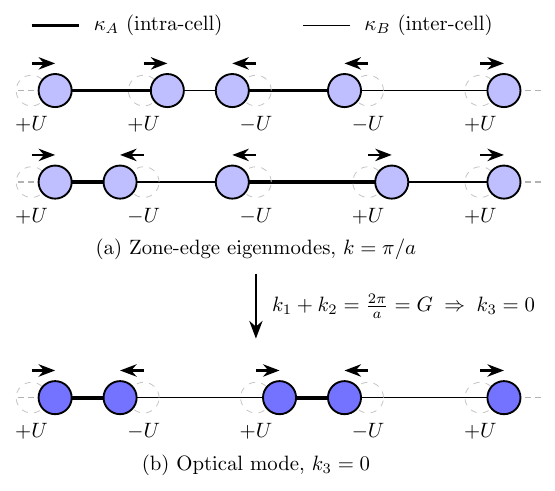}
    \caption{Schematic displacement patterns at play for the first available umklapp process. (a) Acoustic phonons at $k=\pm \pi$ ($a=1$), with $\boldsymbol e_{\pi,-}=(1,1)/\sqrt2$ (since $s(\pi)=1$ for $\Delta\kappa>0$): leading to an  effective displacement $+U,+U,-U,-U,+U,\dots$. The strong (A) springs inside each cell are not strained, whereas every weak (B) spring is strained by $\mp2U$, so that the squared B-spring extension is the same for all cells. (b) The optical $k=0$ eigenmode, $\boldsymbol e_{0,+}=(-1,1)/\sqrt2$: displacements $+U,-U,+U,-U,+U,\dots$, in which A springs are stretched and B springs compressed by the same amount. The uniform force $\propto\gamma_BU^2$ generated by the cubic term on the B springs projects onto (b) (with $D_B(0,+)=-\sqrt2$) and not onto the uniform translation.}
    \label{fig:displacement_patterns}
\end{figure}

Fig. \ref{fig:displacement_patterns} shows a simple scheme of the spatial displacement patterns for the acoustic $k=\pm \pi$ modes fusion and the optical $k=0$ mode. For $\Delta\kappa>0$ there is a single real acoustic standing wave at $k=\pi$ ($\pm\pi$ are the same Bloch state), so the process is the fusion of two phonons of the same mode. Its strain field $d_B\propto(-1)^n$ has a uniform square, $d_B^2\propto$ const, which is a $k=0$ forcing; it couples through $\Gamma_{--+}(\pi,\pi)=-2\sqrt2\,\gamma_B$ (Appendix, Eq.~\eqref{eq:corner_vertex}) to the zone-center optical mode, because $D_A(\pi,-)=0$ and $D_B(\pi,-)=D_B(0,+)=-\sqrt2$. This process can be described as almost rigid dimers vibrating in alternate opposite directions producing a strain in the weak bonds that join the dimers.

\begin{figure}
    \centering
    \includegraphics[width=1.05\linewidth]{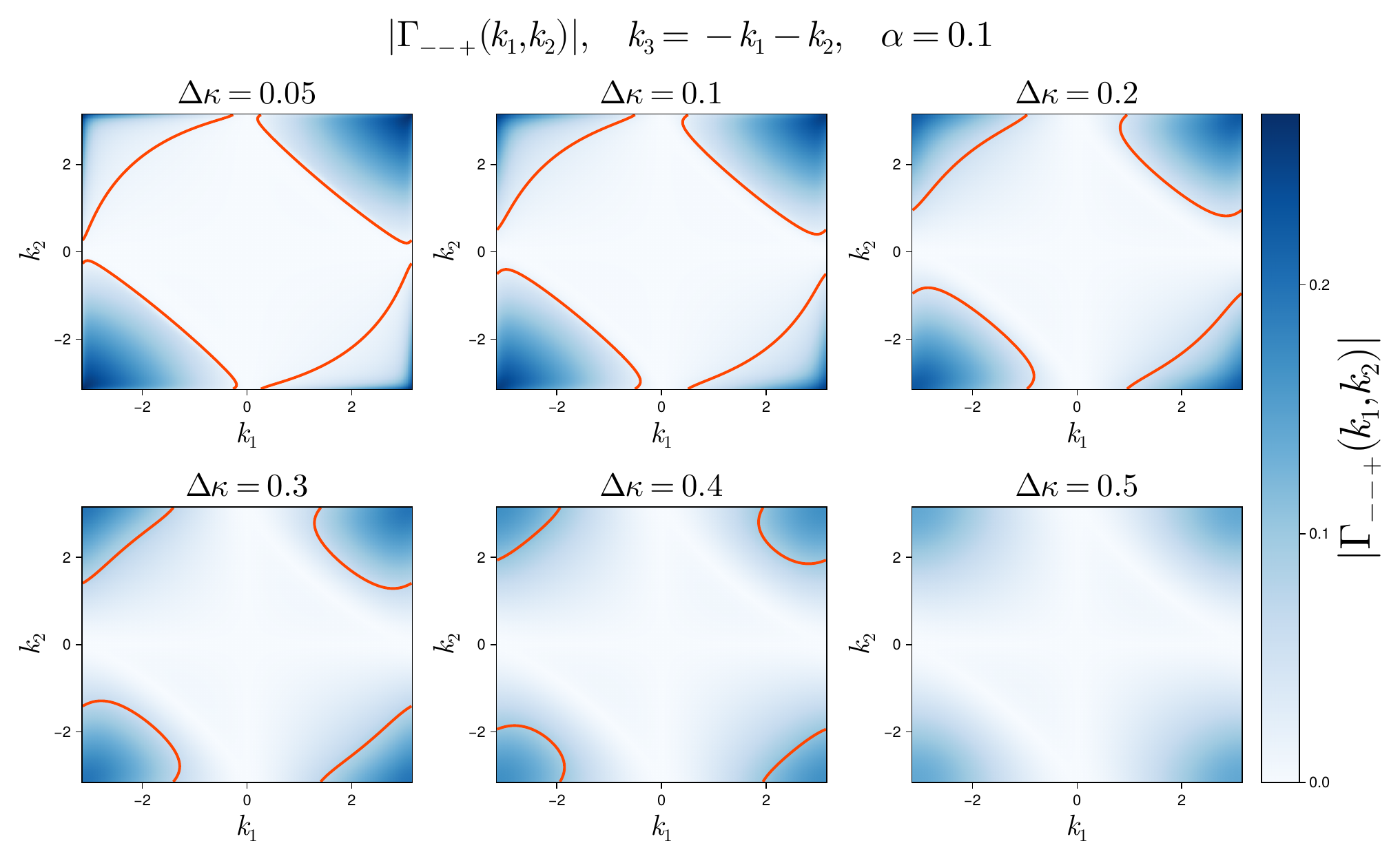}
    \caption{Heatmap  of the three wave vertex coefficient norm $|\Gamma_{-,-,+}(k_1,k_2,k_3)|$ with $k_3=-k_1-k_2$  for different values of $\Delta \kappa$. The red curves are the resonance manifolds for the corresponding $\Delta \kappa$ values. Energy transfer is favored at locations where the red curves touch the darker regions. The integral of such overlap over the whole Brillouin zone divided by the product of phonon frequencies gives the total scattering rate $\mathcal R(\eta)$. }
    \label{fig:overlap_heatmap_resonance_curves_delta_sweep}
\end{figure}

 Now we discuss the second ingredient that comes into play for energy transfer. Once the resonant condition is satisfied, still  the interaction vertex magnitude needs to be taken into account. Fig. \ref{fig:overlap_heatmap_resonance_curves_delta_sweep} shows the situation for several $\Delta \kappa$. The color scale indicates the norm of $|\Gamma_{--+}(k_1,k_2,k_3)|$ and overimposed to it,  we plot in red the resonant manifold. It is clearly seen that in some cases $|\Gamma_{--+}(k_1,k_2,k_3)|\approx 0$ while in principle it is possible to have a resonance due to the momentum and energy conservation. The energy-transfer rate from the acoustic to the optical branch is controlled by the Fermi golden rule, written for the canonical amplitudes \cite{bustamante2019exact,pezzi_three-wave_2021}  $a_k=(\omega_kQ_k+iP_k)/\sqrt{2\omega_k}$, which brings the factor $1/(8\omega_1\omega_2\omega_3)$ in the vertex,

\begin{equation}
W
\propto
\frac{
|\Gamma_{--+}(k_1,k_2,k_3)|^2
}{
8\,\omega_-(k_1)\omega_-(k_2)\omega_+(k_3)
}\,
\frac{\Lambda}{
\left[
\omega_-(k_1)
+
\omega_-(k_2)
-
\omega_+(k_3)
\right]^2
+
\Lambda^2}
.
\end{equation}
The conditions required for an efficient energy transfer are:

\begin{enumerate}
\item the vertex interaction element is large;
\item the resonant phase space is large; and
\item the vertex norm maxima overlap with the resonant manifold over a substantially large part of the phase space.
\end{enumerate}

To include these three conditions at the same time, namely a large vertex element, a large resonant phase space, and their overlap, we define 
\begin{equation}
\mathcal R(\eta)
=
\int dk_1\,dk_2\,
\frac{|\Gamma(k_1,k_2)|^2\delta\!\Bigl(
\omega_-(k_1)
+
\omega_-(k_2)
-
\omega_+(k_3)
\Bigr)}{8\,\omega_-(k_1)\omega_-(k_2)\omega_+(k_3)}
\,
,
\label{eq:scattering_rate}
\end{equation}
with $k_3=-k_1-k_2$ and $\Gamma\equiv\Gamma_{--+}$.

\begin{figure}
    \centering
    \includegraphics[width=1.0\linewidth]{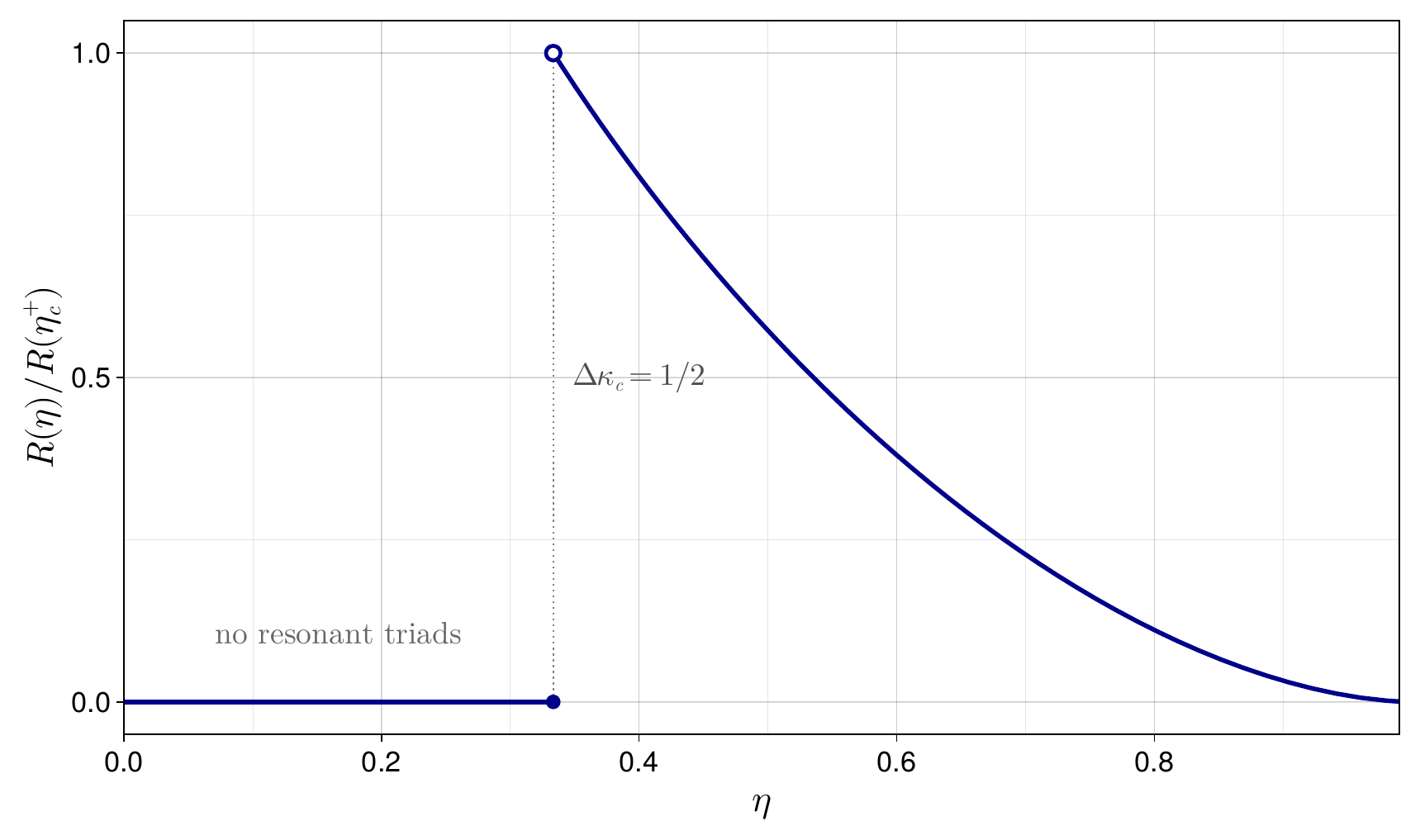}
    \caption{Effective acoustic-to-optical (-,-,+) channel scattering rate $\mathcal R(\eta)$ 
    normalized by its maximum value at $\eta=1/3$. $\mathcal R(\eta)$ vanishes identically for $\eta < \frac{1}{3}$ ($\Delta \kappa > 0.5$), jumps to a finite value at the threshold, is largest just below it ($\Delta \kappa < 0.5$) and decreases toward zero as $\eta \to 1$ ($\Delta \kappa \to 0$).}
    \label{fig:scattering_rate}
\end{figure}

The quantity $\mathcal R(\eta)$ represents the effective acoustic-to-optical scattering rate. Fig  \ref{fig:scattering_rate} presents a plot of $\mathcal R(\eta)$ as a function of $\eta$ showing that there is no energy transfer for $\eta<1/3$, corresponding to $\Delta \kappa > 1/2$.

The behavior of $\mathcal R$ follows from the two ingredients. (i) For $\Delta\kappa>1/2$ the function $F$ is negative everywhere, so $\mathcal R=0$. (ii) As $\Delta\kappa\to1/2^-$ the resonant set is a small closed loop around the corner $(\pi,\pi)$, which is a maximum of $F$; the density of resonant states of a two-dimensional maximum is finite, so $\mathcal R$ does not vanish at the threshold but drops discontinuously to zero on crossing it. There the vertex takes its largest value (Eq.~\eqref{eq:corner_vertex}), so the overlap condition 3 is best satisfied. (iii) As $\Delta\kappa\to0$ the on-shell vertex vanishes linearly in $\Delta\kappa$ (see above), so $\mathcal R\propto\Delta\kappa^{2}$ even though the resonant phase space is largest.

In summary, a three-wave channel exists only for $\Delta \kappa \le 1/2$. Additional relaxation must come from higher-order processes or from quasi-resonances. In fact, as $\Delta \kappa \rightarrow 1$, each frequency dispersion branch flattens. Therefore, the modes in a branch become softer and quasi-resonant. When both effects are combined, the Chirikov criterion \cite{izrailev1966statistical} implies that thermalization is enhanced within each branch. In other words, energy mixes among acoustic modes or among optical modes.

This is supported by  Fig. \ref{fig:entropy_fbc_size}, where $\Delta \kappa =0.1$ relaxes faster than $\Delta \kappa =0.05$, as $\mathcal{R}(\eta)$ grows with $\Delta \kappa$. This is consistent with the fact that only higher-order processes exist in the cubic uniform-chain limit \cite{OnoratoPNAS_2015, malishava_flach_2022}. In the following sections we discuss how the dimerization ratio $\eta$ affects the two main ingredients of the energy transfer. 

\subsection{The vertex amplitude grows with \texorpdfstring{$\eta$}{eta}}

The coupling of an acoustic leg to the strong springs is governed by the bond factor

\begin{equation}
1-s(k),
\end{equation}
whose magnitude increases as $\eta$ grows ($|1-s|\simeq\eta|\sin k|$ for $\eta\ll1$, $|1-s|=2|\sin(k/4)|$ at $\eta=1$). Consequently, the largest value of the interaction element $\Gamma_{--+}$ for the $--+$ channel, which is the main responsible for pumping energy from the acoustic to the optical branch, increases with $\eta$:  it is attained at the corner $k_1=k_2=\pi$ and equals exactly $2\sqrt2\,\gamma_B=2\sqrt2\,\eta\gamma_A$ (Appendix). This is not by itself the on-shell vertex: at $\eta=1$ the vertex vanishes for normal processes by interference, which is why $\mathcal R$ is not monotonic in $\eta$.

\subsection{The phonon gap decreases with \texorpdfstring{$\eta$}{eta}}

For the SSH chain, the optical-acoustic gap is measured at the zone boundary,
$\Delta(\eta)=\omega_+(\pi)-\omega_-(\pi)$. Using the band edges of
Eqs.~\eqref{eq:ac_bounds}--\eqref{eq:opt_bounds} together with
$\Delta\kappa=(1-\eta)/(1+\eta)$, this evaluates to,
\begin{align}
\Delta(\eta)
&=
\frac{2\sqrt{2}\left(1-\eta\,\right)}{\sqrt{1+\eta}(1+\sqrt{\eta})}
\;\simeq\;
\frac{(1-\eta)}{\sqrt{2}},
\qquad \eta \to 1,
\label{eq:gap_vs_eta}
\end{align}
so the gap closes linear in $\eta$ in the uniform-chain limit, and
saturates at $\Delta(0)=2$ in the isolated-dimer limit.

Thus, in the strongly dimerized limit $\eta\to 0$
the system approaches a collection of isolated dimers and the optical-acoustic gap is large. This corresponds to the acoustic branch becoming floppy (flat), and provides a complementary, purely kinematic reason for the suppression of acoustic-acoustic-optical resonances as $\Delta \kappa \to 1$: not only does the gap widen, but the acoustic frequencies feeding the resonance condition (Eq. \ref{eq:quasiresonance_condition}) collapse toward zero, making the term $2\omega_{-}^{\max}$ even less able to reach $\omega_{+}^{\min}$ regardless of the vertex magnitude. In contrast, $\eta\to 1$ corresponds to the uniform-chain limit, where the gap closes and the distinction between acoustic and optical branches becomes progressively weaker. As a result, increasing $\eta$ increases the largest vertex element and simultaneously reduces the energetic separation between the two branches (the gain in resonant phase space being offset near $\eta=1$ by the interference suppression of the on-shell vertex). 

\subsection{Interpretation}

As seen in Fig. \ref{fig:overlap_heatmap_resonance_curves_delta_sweep}, a particularly important feature of the SSH model is that, except near the threshold, the location of the largest interaction vertex does not coincide with the location of the resonant manifold. In fact, as shown in the appendix,  the vertex of the $--+$ channel is largest at the zone corners $k_1=k_2=\pm\pi$ for \emph{all} $\eta$: for $\eta\ll1$ it is smooth, $|\Gamma_{--+}|\simeq2\sqrt2\,\gamma_B\,|\sin\frac{k_1}{2}\sin\frac{k_2}{2}\cos\frac{k_1+k_2}{2}|$, with maximum $2\sqrt2\gamma_B$ at the corners (the ridges at $k=\pm\pi/2$ coming from the strong-bond term are $O(\eta^2)$ and subdominant); for $\eta\to1$ it is $|\Gamma_{--+}|^2=32\gamma_A^2\sin^2\frac{k_1}{4}\sin^2\frac{k_2}{4}\sin^2\frac{k_1+k_2}{4}$ in the umklapp region and vanishes in the normal region. The resonant manifold of Fig.~\ref{fig:resonance_level_set_alpha} shrinks onto this corner as $\Delta\kappa\to1/2^-$, which is why the overlap is largest just below threshold, whereas at small $\Delta\kappa$ the manifold runs through the region where the vertex is small.

As shown in the appendix, for $\eta=1$ the A- and B-bond contributions to $\Gamma_{--+}$ interfere destructively for every normal process ($G=0$), $\Gamma_{--+}\propto\sin(K/4)=0$; physically, the true atomic momentum $q=k/2$ is conserved modulo $2\pi$, so the doubled-cell momentum conservation is too permissive. Numerically, the on-shell vertex goes to zero as $\Delta\kappa\to0$. Hence three-wave processes are ineffective for $\Delta \kappa=0$.  In fact, it is well known that in the monoatomic cubic chain (the $\alpha$-FPUT chain), the dominant energy transfer mechanism is based on higher-order (six-wave) resonant processes \cite{OnoratoPNAS_2015}, whereas in the alternating-masses case, energy transfer is ruled by a three-wave resonant process   \cite{onorato2023wave, pezzi_multi-wave_2025}. As three-wave resonant processes are not effective there, the transference of energy between modes is due to higher-order processes, which are slower than the $3-$wave processes \cite{OnoratoPNAS_2015, bustamante2019exact}. 

\begin{figure}[t]
    \centering
    \includegraphics[width=\linewidth]{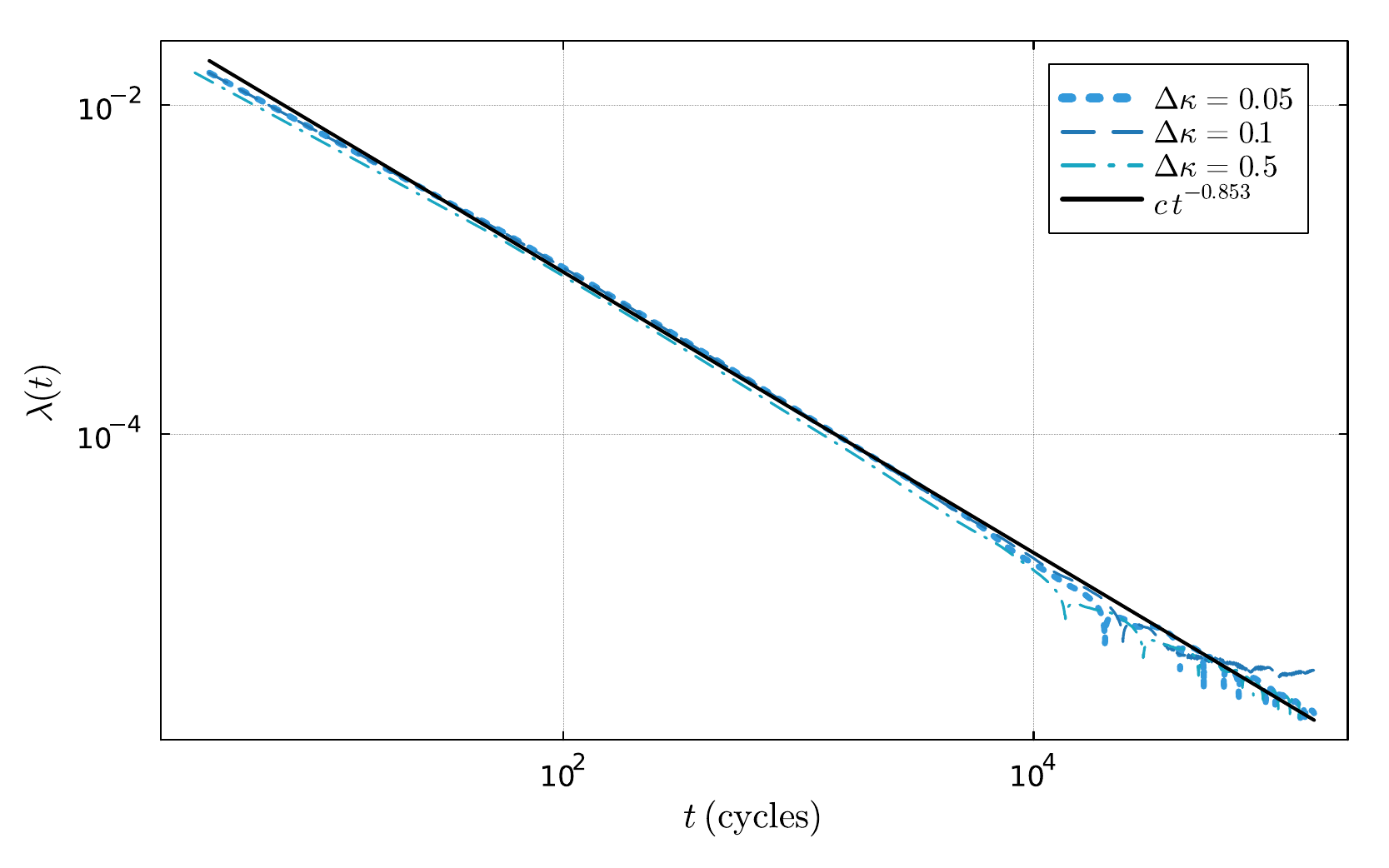}
    \vspace{-1.5em}
    \includegraphics[width=\linewidth]{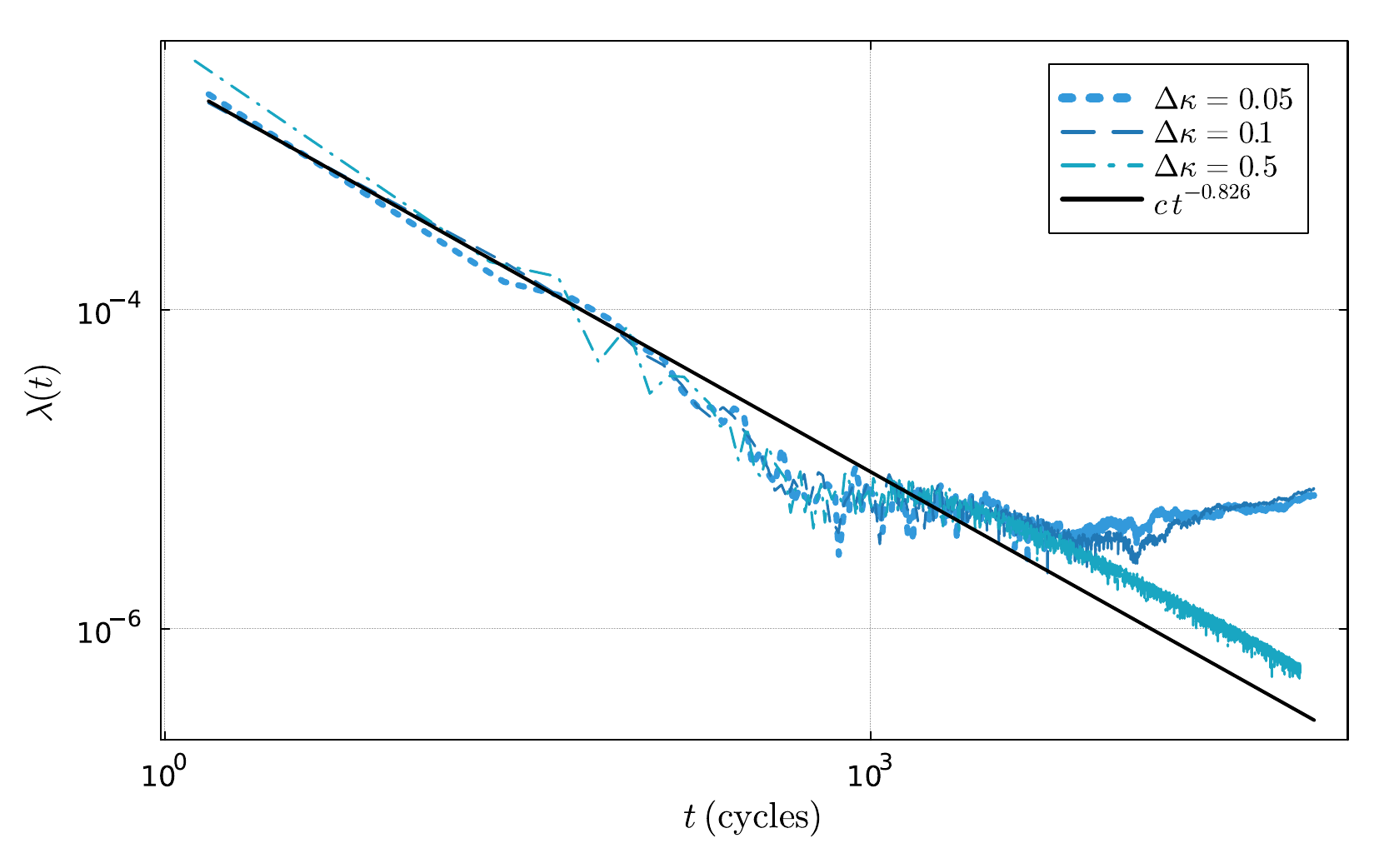}
    \caption{Finite-time maximum Lyapunov exponent $\lambda(t)$ for
    $\alpha = 0.1$, $N = 64$, $\epsilon = 0.0069$.
    \textit{Top}: PBC, second-mode excitation. All three $\Delta\kappa$
    values collapse onto $ct^{-0.853}$, showing that the gap leaves the
    local phase-space geometry unchanged.
    \textit{Bottom}: FBC, first-mode excitation. Curves separate near
    $t \sim 10^3$: $\Delta\kappa = 0.05$ and $0.1$ depart from the power-law decay, while $\Delta\kappa = 0.5$ continues to decay as $ct^{-0.826}$, consistent with the delayed transfer in the three-wave threshold.}
    \label{fig:ftmle_pbc_fbc}
\end{figure}

We end up saying that while the present discussion explains some general features of thermalization seen in the numerical simulations, still there are aspects of the problem that are very intriguing. As seen in Fig.  \ref{fig:SpectralEntropy_Different_Springs_Fixed_Conditions_N64_alpha_second_mode}, we show that for FBC exciting the second mode, the system is  trapped as happens for the second mode with PBC. Only for $\Delta\kappa 0.1$ and $0.2$, these mode excitations lead to substantial entropy growth. This seems to be a case of a sticky state, as the ones found in a previous work for the $N=3$ case \cite{toledo2018escape}. The finite-time Lyapunov exponent (Fig.~\ref{fig:ftmle_pbc_fbc}) provides complementary evidence for this picture: under FBC the exponent separates with respect to a power-law behavior, while under PBC all $\Delta \kappa$ values collapse onto the same decay, indicating that the gap controls global energy transfer rather than the local divergence between trajectories. 

\section{Conclusions}\label{sec:conclusions}

The nonlinear relaxation and spectral entropy in FPUT chains with alternating spring constants were studied as a function of the dimerization strength $\Delta\kappa$, the $\alpha$-type nonlinearity, and boundary conditions. 
The central finding was that the thermalization in the FPUT diatomic chain is delayed as the dimerization increases. The interbranch energy transfer  is governed by two effects. One is the resonant manifold, controlled by the phononic gap width and momentum conservation. The other is the magnitude of the three wave vertex coefficient for the acoustic-acoustic to optical channel. Therefore, the introduction of $\alpha$ nonlinearities admitted three-wave resonant processes: according to the wave turbulence framework \cite{l2010discrete, nazarenko_wave_2011, lvov_double_2018, pezzi_multi-wave_2025}, interbranch energy transfer was driven by resonant triads for which frequencies from the acoustic branch summed to match a target optical frequency. A short summary of results is the following:

\begin{enumerate}
    \item The dispersion relation gap acted as a tunable barrier to delay interbranch energy transfer. Increasing $\Delta\kappa$ widened the gap and progressively modifies the transfer into the optical branch, and ultimately, controlling thermalization time, which is delayed above $\Delta\kappa=1/2$ for all system sizes ($N\le256$) and effective energy densities studied. Nevertheless, at long times in FBC ($10^7$ periods), the closed channels also approach equipartition, and the delay generated by the three-wave resonances decreases as $N$ grows. This slow transfer observed above the threshold at the largest $N$ and nonlinearity is consistent with higher-order processes  not blocked by the three-wave threshold

    \item  Three wave resonances acoustic to optical are not possible for $\Delta \kappa  > 1/2$, i.e., approaching the dimer liquid limit region. The resonant manifold opens for $\Delta\kappa<1/2$ as a small loop around the zone corner, corresponding to a collision of two acoustical stationary phonons that generate the highest possible frequency optical phonon which arises at the center of the Brillouin zone. At the linear-chain limit $\Delta \kappa = 0$ the resonant set exists kinematically but the three-wave vertex vanishes on it by A/B-bond interference, consistent with the well-known dominance of higher-order processes in the monoatomic cubic chain \cite{OnoratoPNAS_2015, malishava_flach_2022}.
        
    \item The optical branch activation was controlled by accessible resonance conditions and by the norm of the vertex. The resonance manifold changed its location as $\Delta \kappa$ increased and the largest vertex, $2\sqrt2\,\gamma_B$, decreased as the system went from the linear chain to a liquid of nonlinear dimers.

    \item The vertex of the optical-pumping channel is maximal at the zone corners $k_1=k_2=\pi$ for all $\eta$. The scattering rate $\mathcal R$ is non-monotonic in $\Delta\kappa$: it grows as $\Delta\kappa^2$ at small dimerization, is largest just below $\Delta\kappa=1/2$ where the resonant loop collapses onto the vertex maximum, and drops abruptly to zero at the threshold.  Phonon diffusion also decreases as the chain is dimerized.

    \item However, the entropy and relaxation is far from trivial as there is a competing effect due to the band flattening,i.e., the optical and acoustic bands become flat for $\Delta \kappa \rightarrow 1$ and thus almost all acoustic and optical modes become quasi-resonant, speeding up the relaxation inside dispersion branches according to the Chirikov criteria \cite{izrailev1966statistical}. As no exact three-wave channel exists above $\Delta \kappa \ge 1/2$,  the remaining transfer comes from higher-order processes.
        
    \item Boundary and initial conditions modulated the effective resonance order: FBC imposed higher-order requirements on interbranch coupling by lowering the acoustic frequency ladder,  yet above the three-wave resonance threshold the system relaxes completely, though later, within $10^7$ periods, while PBC instead trapped most $\Delta\kappa$ values in long-lived states, with fast entropy growth occurring only in a narrow window ($\Delta \kappa \approx 0.1$-$0.2$).
    
\end{enumerate}

More broadly, our results illustrate how the SSH-like spectral gap  — inherited here from the same chiral bulk symmetry exploited for edge-state protection in related dimerized lattices \cite{zak1989berry, berg2010bulk, susstrunk2016classification, many2022nonlinear, sone2025} — can be made to interact non-trivially with the nonlinear coupling itself when both descend from the same microscopic potential, offering a route to engineer nonlinear energy transport in gapped lattices. 

\section*{Data and Code Availability}\label{sec:data_and_code}

The custom Julia code used for all simulations and postprocessing is available in a public repository at \href{https://github.com/ak3sit0/FPUT-diatomic-chain}{https://github.com/ak3sit0/FPUT-diatomic-chain}; a detailed README file with installation and usage instructions is included. The numerical  data generated in this study (modal energy histories, spectral entropy time series, and raw integration output for all parameter sets) are available from the corresponding authors upon reasonable request.

\section*{Acknowledgments}

The authors acknowledge and express their gratitude to Carlos Ernesto López Natarén for assistance with the high-performance computing infrastructure at the Instituto de Física, UNAM, where we ran our calculations, and for his valuable support. This work was supported by UNAM DGAPA PAPIIT project IN101924. José A. Aké is supported by a CONAHCyT PhD scholarship.

\textbf{Correspondence and requests for materials} should be addressed to José A. Aké (\href{mailto:akejja@estudiantes.fisica.unam.mx}{akejja@estudiantes.fisica.unam.mx}) and Gerardo Naumis (\href{mailto:naumis@fisica.unam.mx}{naumis@fisica.unam.mx}).

\section{Appendix}\label{sec:appendix}

\subsection{Coupling coefficients}
   Let
\[
s_j := \frac{\mathcal{S}(k_j)}{|\mathcal{S}(k_j)|}= \frac{1+\eta e^{-ik_j}}{\sqrt{1+\eta^2+2\eta\cos k_j}},
\qquad
\zeta_j = e^{ik_j}s_j,
\]
and $\sigma_j=+1$ for the optical branch and $\sigma_j=-1$ for the acoustic branch. With $D_A=(1+\sigma s)/\sqrt2$ and $D_B=-(1+\sigma\zeta)/\sqrt2$,
\begin{align}
\Gamma^{A}
&=
\frac{1}{2^{3/2}}
\prod_{j=1}^{3}
\left(1+\sigma_j s_j\right),\\
\Gamma^{B}
&=
-\frac{1}{2^{3/2}}
\prod_{j=1}^{3}
\left(1+\sigma_j e^{ik_j}s_j\right),
\end{align}
and $\Gamma_{\sigma_1\sigma_2\sigma_3}=\gamma_A(\Gamma^{A}+\eta\,\Gamma^{B})$. Using the momentum conservation $k_1+k_2+k_3=0$ and restoring the arguments of the functions we obtain,
\begin{align*}
\Gamma^{A}_{\sigma_1 \sigma_2 \sigma_3}(k_1,k_2)
&=
\frac{1}{2^{3/2}}
\Bigl[1+\sigma_1 s(k_1)\Bigr]
\Bigl[1+\sigma_2 s(k_2)\Bigr] \notag\\
&\quad \times
\Bigl[1+\sigma_3 s(-k_1-k_2)\Bigr],
\end{align*}
and, 
\begin{align*}
\Gamma^{B}_{\sigma_1 \sigma_2 \sigma_3}(k_1,k_2)
&=
\frac{-1}{2^{3/2}}
\Bigl[1+e^{ik_1}\sigma_1 s(k_1)\Bigr]
\Bigl[1+ e^{ik_2}\sigma_2s(k_2)\Bigr] \notag\\
&\quad \times
\Bigl[1+ e^{-i(k_1+k_2)}\sigma_3s(-k_1-k_2)\Bigr].
\end{align*}
At the zone center $s(0)=1$ and the boundary $s(\pi)=1$. In particular, for the channel $--+$ at the corner $k_1=k_2=\pi$, $k_3=-2\pi\equiv0$: $1-s(\pi)=0$ cancels $\Gamma^A$ exactly, while all three factors of $\Gamma^B$ equal $2$, so that, \emph{for every $\eta$},
\begin{equation}
\Gamma_{--+}(\pi,\pi)=-2\sqrt2\,\gamma_B,\qquad |\Gamma_{--+}(\pi,\pi)|=2\sqrt2\,\eta\,\gamma_A .
\label{eq:corner_vertex}
\end{equation}
A numerical scan of the whole $(k_1,k_2)$ zone shows that this is the global maximum of $|\Gamma_{--+}|$ for all $\eta$ tested ($0.05\le\eta\le0.99$).

\subsection{Symmetry analysis of the full three-phonon vertex}

The squared coupling vertex can be written as

\begin{equation}
|\Gamma|^2
=
\gamma_A^2
\left[
|\Gamma^{A}|^2
+
2\eta\,\mathrm{Re}
\!\left(
\Gamma^{A}(\Gamma^{B})^{*}
\right)
+
\eta^2 |\Gamma^{B}|^2
\right],
\end{equation}
with

\begin{equation}
s_j=s(k_j),
\qquad
k_3=-(k_1+k_2).
\end{equation}

We now analyze the symmetries of $|\Gamma(k_1,k_2)|^2$.

\subsubsection{Exchange symmetry}

Consider first the case in which the first two branches are identical,

\begin{equation}
\sigma_1=\sigma_2.
\end{equation}

Then

\begin{equation}
\Gamma^{A}(k_1,k_2)
=
\Gamma^{A}(k_2,k_1),
\end{equation}

since

\begin{equation}
(1+\sigma_1 s_1)(1+\sigma_2 s_2)
=
(1+\sigma_1 s_2)(1+\sigma_2 s_1)
\end{equation}
when $\sigma_1=\sigma_2$, and the third factor depends only on $k_1+k_2$. Likewise,

\begin{equation}
\Gamma^{B}(k_1,k_2)
=
\Gamma^{B}(k_2,k_1).
\end{equation}

Therefore,

\begin{equation}
|\Gamma(k_1,k_2)|^2
=
|\Gamma(k_2,k_1)|^2.
\end{equation}

Hence the channels

\begin{equation}
--+,\qquad ++-,\qquad +++,\qquad ---
\end{equation}

are symmetric under reflection about the diagonal

\begin{equation}
k_1=k_2.
\end{equation}
Figs.
 \ref{fig:gamma_dk01} and \ref{fig:gamma_dk03}  presents several cases, showing that indeed such symmetry is observed.
For channels satisfying $\sigma_1\neq\sigma_2$, namely

\begin{equation}
+-+,\qquad -++,\qquad +--,\qquad -+-,
\end{equation}

the exchange symmetry is lost and, in general,

\begin{equation}
|\Gamma(k_1,k_2)|^2
\neq
|\Gamma(k_2,k_1)|^2.
\end{equation}

Instead, these channels occur in mirror pairs,

\begin{equation}
|\Gamma_{+-+}(k_1,k_2)|^{2}
=
|\Gamma_{-++}(k_2,k_1)|^{2},
\end{equation}

and

\begin{equation}
|\Gamma_{+--}(k_1,k_2)|^{2}
=
|\Gamma_{-+-}(k_2,k_1)|^{2}.
\end{equation}

\subsubsection{Dependence on \texorpdfstring{$k_1-k_2$}{k1-k2} and \texorpdfstring{$k_1+k_2$}{k1+k2}}

Introducing

\begin{equation}
u=k_1-k_2,
\qquad
v=k_1+k_2,
\end{equation}

the momentum conservation condition becomes

\begin{equation}
k_3=-v.
\end{equation}

The vertex depends on

\begin{equation}
s\!\left(\frac{u+v}{2}\right),
\qquad
s\!\left(\frac{v-u}{2}\right),
\qquad
s(v),
\end{equation}
The third factor, and the whole B-bond phase $e^{ik_3}=e^{-i(k_1+k_2)}$, depend on $v$ only, so the vertex generically varies with $v=k_1+k_2$ at fixed $u$ and develops structure along lines $k_1+k_2=\mathrm{const}$; we do not find a symmetry of $|\Gamma|^2$ associated with constant relative momentum $u$. The B-bond contribution contains the phase factor

\begin{equation}
e^{ik_3}
=
e^{-i(k_1+k_2)},
\end{equation}

which generates harmonics such as

\begin{equation}
\cos(k_1+k_2),
\qquad
\cos\!\bigl[2(k_1+k_2)\bigr],
\end{equation}

as well as mixed combinations,

\begin{equation}
\cos(k_1+2k_2),
\qquad
\cos(2k_1+k_2).
\end{equation}

As a result, the full vertex tends to develop structures aligned along

\begin{equation}
k_1+k_2=\mathrm{const},
\end{equation}

which explains the ridge-like features observed in the numerical maps in Fig.  \ref{fig:gamma_dk01}.

\subsubsection{Inversion symmetry}

Since

\begin{equation}
s(-k)=s^{*}(k),
\end{equation}

one finds

\begin{equation}
\Gamma^{A}(-k_1,-k_2)
=
(\Gamma^{A}(k_1,k_2))^{*},
\end{equation}

and

\begin{equation}
\Gamma^{B}(-k_1,-k_2)
=
(\Gamma^{B}(k_1,k_2))^{*}
\end{equation}
for $\Gamma^B$ use $e^{-ik}s(-k)=(e^{ik}s(k))^*$; the branch indices $\sigma_j$ are unchanged). Therefore,

\begin{equation}
|\Gamma(-k_1,-k_2)|^2
=
|\Gamma(k_1,k_2)|^2.
\end{equation}

Hence the full vertex always possesses inversion symmetry about the origin of the $(k_1,k_2)$ plane as seen in Figs.
 \ref{fig:gamma_dk01} and \ref{fig:gamma_dk03} .

\subsection{Reduction using real displacements}

These coefficients are in general complex; for real displacements one uses momentum-conserving combinations with \(Q_{-k,\sigma}=Q_{k,\sigma}^*\).
At symmetric points (e.g., \(k_j=0,\pi\)) one has \(s_j=+1\) (for $\Delta\kappa>0$) and many terms simplify or vanish. For real displacements the modal amplitudes satisfy
\[
Q_{-k,\sigma}=Q_{k,\sigma}^* .
\]
Starting from the cubic potential in modal coordinates (unit-cell count \(M=N/2\)) and by 
using the momentum constraint to eliminate $k_3\equiv-(k_1+k_2)$, then
\begin{align*}
V_3 &= \frac{1}{3\sqrt{M}}\sum_{k_1,k_2}
\sum_{\sigma_1,\sigma_2,\sigma_3}
\Gamma_{\sigma_1\sigma_2\sigma_3}\bigl(k_1,k_2,-k_1-k_2\bigr)\, \\
&\quad Q_{k_1,\sigma_1}Q_{k_2,\sigma_2}Q_{-k_1-k_2,\sigma_3}.
\end{align*}

\begin{figure}[t]
    \centering
    \includegraphics[width=1.0\linewidth]{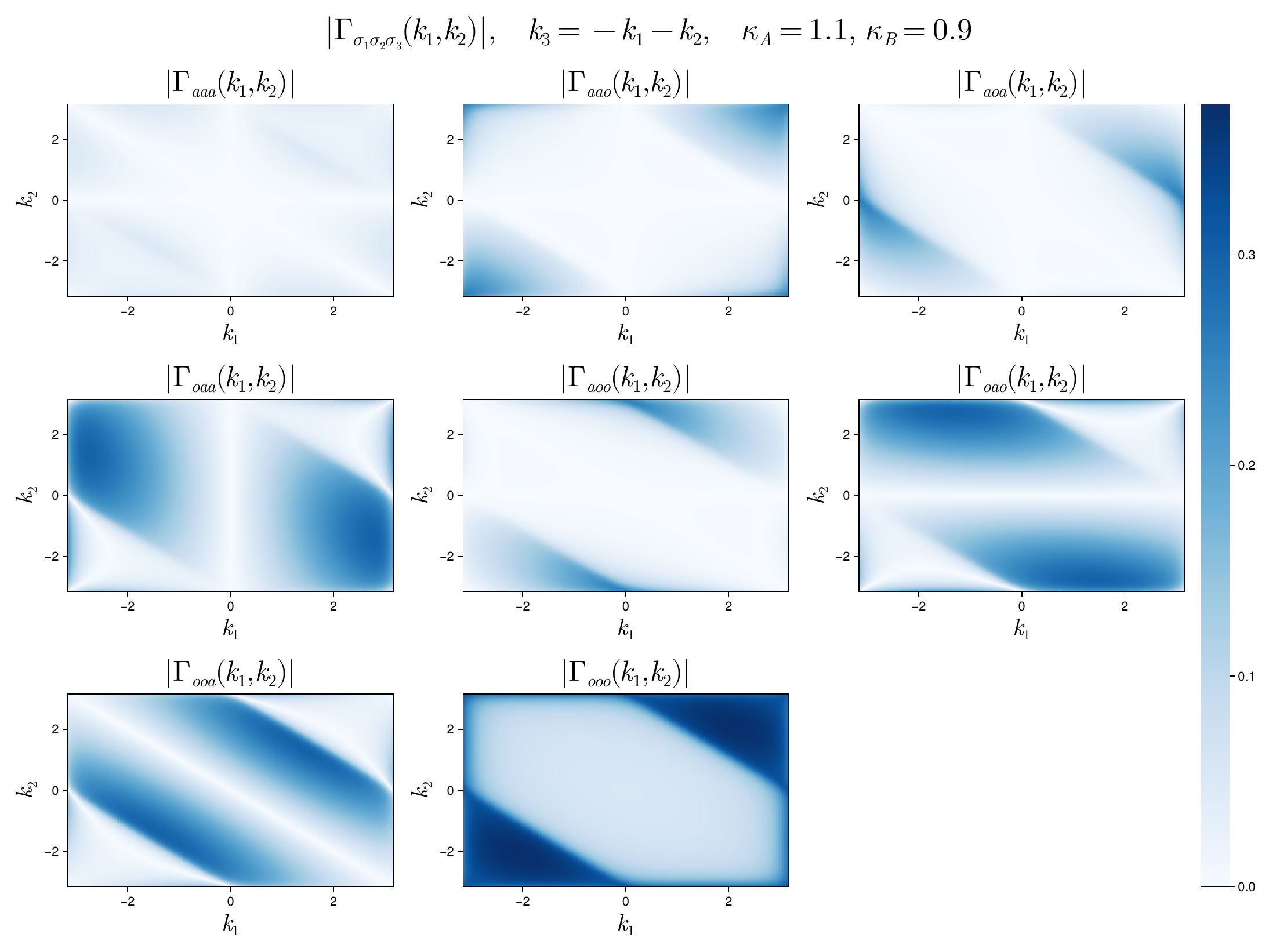}
    \caption{Heatmap showing the magnitude of the coupling coefficients $|\Gamma_{\sigma_1 \sigma_2 \sigma_3}(k_{1}, k_{2})|$ for a fixed $\Delta \kappa = 0.1$ value and $\alpha = 0.1$. Each panel corresponds to one of the eight branch-index combinations $\sigma_1=\pm,\sigma_2=\pm,\sigma_3=\pm$; the third wavevector $k_{3}$ satisfies $k_{3} = -k_{1} - k_{2}$. The color scale is shared across all panels.} 
    \label{fig:gamma_dk01}
\end{figure}

\begin{figure}[t]
    \centering
    \includegraphics[width=1.0\linewidth]{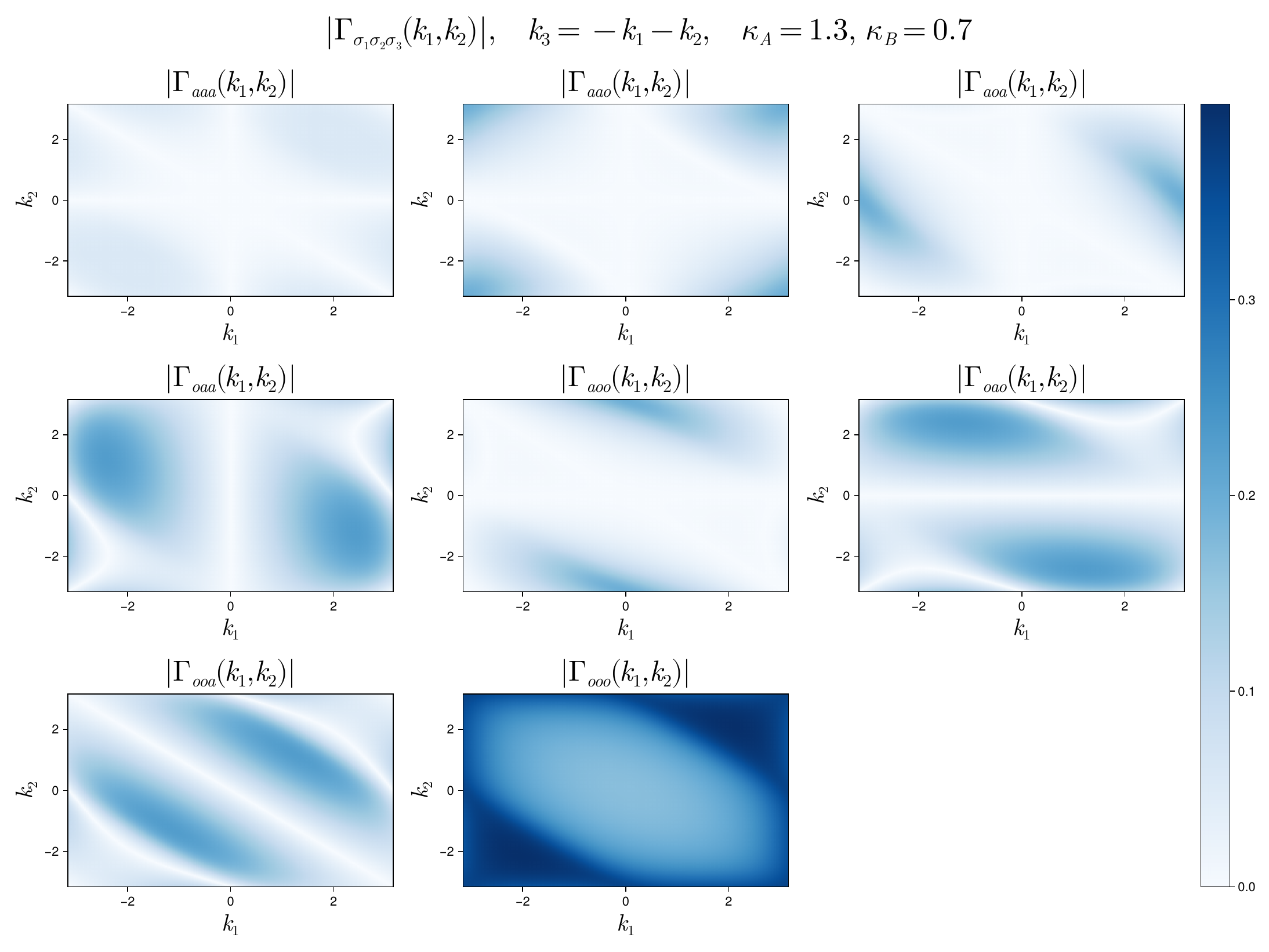}
    \caption{Same as Fig. \ref{fig:gamma_dk01} for $\Delta\kappa = 0.3$}
    \label{fig:gamma_dk03}
\end{figure}

Using the reality condition
\[
Q_{-k,\sigma} = Q_{k,\sigma}^*\quad(\text{same }\sigma),
\]
and the property (which follows from $s(-k)=s^*(k)$),
\[
\Gamma_{\sigma_1\sigma_2\sigma_3}(-k_1,-k_2,-k_3)
= \Gamma_{\sigma_1\sigma_2\sigma_3}(k_1,k_2,k_3)^*,
\]
the term $(-k_1,-k_2)$ is the complex conjugate of the term $(k_1,k_2)$. Because $V_3$ is real, the full sum equals its real part, so one convenient real form is (the sum still running over all discrete momenta, with no extra factor of 2).
\begin{align*}
V_3 &= \frac{1}{3\sqrt{M}}\sum_{k_1,k_2}
\sum_{\sigma_1,\sigma_2,\sigma_3}
\Re\!\Big\{ \Gamma_{\sigma_1\sigma_2\sigma_3}(k_1,k_2,-k_1-k_2)\, \\
&\quad Q_{k_1,\sigma_1}Q_{k_2,\sigma_2}Q_{-k_1-k_2,\sigma_3}\Big\}.
\end{align*}
Figs.
 \ref{fig:gamma_dk01} and \ref{fig:gamma_dk03}  present a color map of $\Gamma_{\sigma_1\sigma_2\sigma_3}(k_1,k_2,-k_1-k_2)$ as a function of $k_1$ and $k_2$.

\subsection{Dimerized limit \texorpdfstring{($\eta \ll 1$)}{(eta << 1)}}

In the dimerized limit, the chain approaches a collection of weakly coupled dimers. Up to
first order in $\eta$,
\begin{equation}
s(k)\simeq1-i\eta\sin k .
\end{equation}
The strong-bond factors are therefore of two kinds: for the \emph{acoustic} legs
\begin{equation}
1-s(k)\simeq i\eta\sin k=O(\eta),
\end{equation}
while for the optical legs $1+s(k)\simeq2$. Each acoustic leg of $\Gamma^{A}$ therefore costs a factor $\eta$: on the strong bond a dimer moving in an acoustic mode is almost rigid. The only channel that survives at zeroth order is the one with three optical legs,
\begin{equation}
\Gamma_{+++}\simeq\gamma_A\,\frac{2^{3}}{2^{3/2}}=2\sqrt2\,\gamma_A ,
\end{equation}
which is momentum independent. For the optical-pumping channel the strong bond gives only
\begin{equation}
\Gamma^{A}_{--+}\simeq-\frac{\eta^{2}}{\sqrt2}\,\sin k_1\sin k_2 ,
\end{equation}
of order $\eta^{2}$. The weak bond, instead, contributes $\eta\,\Gamma^{B}$ with factors that remain of order one: with $\zeta=e^{ik}s\simeq e^{ik}$,
\begin{equation}
1-e^{ik}\simeq-2i\,e^{ik/2}\sin\frac{k}{2},
\qquad
1+e^{ik}\simeq2\,e^{ik/2}\cos\frac{k}{2}.
\end{equation}
For channels with two or more acoustic legs this term dominates, and for $k_3=-k_1-k_2$
\begin{equation}
\Gamma_{--+}\simeq\eta\,\gamma_A\,\Gamma^{B}_{--+}
=2\sqrt2\,\gamma_B\,
\sin\frac{k_1}{2}\,\sin\frac{k_2}{2}\,\cos\frac{k_1+k_2}{2},
\end{equation}
which is maximal, $|\Gamma_{--+}|\simeq2\sqrt2\,\gamma_B$, at the zone corners $k_1=k_2=\pm\pi$ (in agreement with the exact corner value, Eq.~\eqref{eq:corner_vertex}). The ridges of $\Gamma^{A}_{--+}$ at $|\sin k|=1$ ($k=\pm\pi/2$) are thus subdominant, and the vertex of the optical-pumping channel vanishes linearly in $\eta$. The strongly dimerized regime $\eta<1/3$ lies entirely above the three-wave threshold; this limit characterizes the vertex, not an open channel. Channels with one acoustic leg receive comparable $O(\eta)$ contributions from both bonds.

\subsection{Uniform-chain limit \texorpdfstring{($\eta \to 1$)}{(eta -> 1)}}
We make an expansion around $\eta=1$ using a small parameter,
\begin{equation}
\eta=1-\delta,
\qquad
\delta\ll1.
\end{equation}
At $\eta=1$, for $|k|<\pi$,
\begin{equation}
s(k)
=
\frac{1+e^{-ik}}
{\sqrt{2+2\cos k}}
=
e^{-ik/2},
\end{equation}
(the limit is non-uniform in a layer of width $O(\delta)$ around $k=\pm\pi$, where $s(\pm\pi)=1$ for $\eta<1$ but $s\to\mp i$ at $\eta=1$). Therefore
\begin{equation}
1-s(k)
=
1-e^{-ik/2}
=
2i\,e^{-ik/4}
\sin\frac{k}{4},
\end{equation}
and
\begin{equation}
1+s(k)
=
2e^{-ik/4}
\cos\frac{k}{4}.
\end{equation}
Consequently,
\begin{equation}
|1-s(k)|^2
=
4\sin^2\frac{k}{4},
\qquad
|1+s(k)|^2
=
4\cos^2\frac{k}{4}.
\end{equation}
\subsubsection{Vertex maxima}

Unlike the dimerized limit, there is no small parameter suppressing acoustic channels. Instead, the vertex is controlled by the competition between
$\sin\frac{k}{4}$ and $\cos\frac{k}{4}$. The acoustic factor is largest at the Brillouin-zone boundary,
\begin{equation}
k=\pm\pi.
\end{equation}
Indeed,
\begin{equation}
|1-s(\pi)|^2=2,
\qquad
|1+s(\pi)|^2=2,
\end{equation}
while
\begin{equation}
|1-s(0)|^2=0.
\end{equation}
Thus acoustic factors vanish at $k=0$ and are maximal near $k=\pi$.

\subsubsection{General structure and interference}
For the A bond, a leg with $\sigma=+$ contributes $2\cos\frac k4\,e^{-ik/4}$ and a leg with $\sigma=-$ contributes $2i\sin\frac k4\,e^{-ik/4}$; for the B bond, $1+\sigma e^{ik}s=1+\sigma e^{ik/2}$ gives $2\cos\frac k4\,e^{ik/4}$ and $-2i\sin\frac k4\,e^{ik/4}$, respectively. With $n_a$ the number of acoustic legs, 
\begin{equation}
K\equiv k_1+k_2+k_3=2\pi G
\end{equation} 
(all $k_j$ reduced to $(-\pi,\pi)$), and
\begin{equation}
C\equiv\prod_{\sigma_j=+}\cos\frac{k_j}{4}\prod_{\sigma_j=-}\sin\frac{k_j}{4},
\end{equation}
one finds, using $\gamma_A=\gamma_B$ at $\eta=1$,
\begin{equation}
\Gamma_{\sigma_1\sigma_2\sigma_3}=2\sqrt2\,\gamma_A\,C\left[i^{n_a}e^{-iK/4}-(-i)^{n_a}e^{iK/4}\right].
\end{equation}
For an odd number of optical legs, the number of possible acoustic legs is $n_a=0,2$, thus,
\begin{equation}
|\Gamma|=4\sqrt2\,\gamma_A\,|C||\sin(K/4)|
\end{equation}
while for an even number of optical legs ($n_a=1,3$),
\begin{equation}
|\Gamma|=4\sqrt2\,\gamma_A\,|C||\cos(K/4)|, 
\end{equation}
Hence, at $\eta=1$, when the number of optical legs is odd, only Umklapp processes ($G=\pm1$, $|\sin(K/4)|=1$) survive and normal ones ($G=0$) vanish identically; when it is even only normal processes survive. This is the statement that the true atomic momentum is conserved modulo $2\pi$.

\subsubsection{Single-optical channel \texorpdfstring{$--+$}{--+}}
For the most important optical pumping channel, $(\sigma_1,\sigma_2,\sigma_3)=(-,-,+)$,
\begin{equation}
\Gamma^{A}_{--+}=-2\sqrt2\;e^{-iK/4}\,
\sin\frac{k_1}{4}\sin\frac{k_2}{4}\cos\frac{k_3}{4},
\end{equation}
\begin{equation}
\Gamma^{B}_{--+}=2\sqrt2\;e^{iK/4}\,
\sin\frac{k_1}{4}\sin\frac{k_2}{4}\cos\frac{k_3}{4}.
\end{equation}
Both terms have the same modulus and differ only in the phase $e^{\pm iK/4}$. Since $\gamma_A=\gamma_B$ at $\eta=1$, both must be kept, and their sum is
\begin{equation}
    \begin{split}
\Gamma_{--+}=\gamma_A\left(\Gamma^{A}_{--+}+\Gamma^{B}_{--+}\right)
=\\4\sqrt2\,i\,\gamma_A\,\sin\frac{K}{4}\,
\sin\frac{k_1}{4}\sin\frac{k_2}{4}\cos\frac{k_3}{4}.
\end{split}
\end{equation}

Momentum conservation fixes $K=2\pi G$, with $k_3\in(-\pi,\pi)$. For normal processes ($G=0$) the interference factor $\sin(K/4)$ vanishes and $\Gamma_{--+}=0$ identically. For Umklapp processes ($G=\pm1$, i.e., $|k_1+k_2|>\pi$) one has $|\sin(K/4)|=1$ and $|\cos(k_3/4)|=|\sin\frac{k_1+k_2}{4}|$, so that
\begin{equation}
|\Gamma_{--+}|^{2}
=32\,\gamma_A^{2}\,
\sin^{2}\frac{k_1}{4}\,\sin^{2}\frac{k_2}{4}\,\sin^{2}\frac{k_1+k_2}{4},
\qquad |k_1+k_2|>\pi .
\end{equation}
The maximum, $|\Gamma_{--+}|=2\sqrt2\,\gamma_A$, is reached at the zone corners $k_1=k_2=\pm\pi$. (We checked this expression numerically against the full vertex at $\eta=1-10^{-9}$.)
In both limits the maximum of $|\Gamma_{--+}|$ lies at the zone corners $k_1=k_2=\pm\pi$.

\subsubsection{Other channels at \texorpdfstring{$\eta=1$}{eta=1}}

The remaining channels follow from the general formula. For $(\sigma_1,\sigma_2,\sigma_3)=(+,-,-)$ (one optical leg, Umklapp only; the optical leg carries $\cos$, the acoustic legs $\sin$),
\begin{equation}
|\Gamma_{+--}|^2
=
32\,\gamma_A^2\,
\cos^2\frac{k_1}{4}
\sin^2\frac{k_2}{4}
\sin^2\frac{k_3}{4}\;\sin^2\frac{K}{4},
\end{equation}
and similarly $|\Gamma_{-+-}|^2=32\gamma_A^2\sin^2\frac{k_1}{4}\cos^2\frac{k_2}{4}\sin^2\frac{k_3}{4}\sin^2\frac K4$. The weight is therefore enhanced for $k_1\to0$ (optical leg at zone center) and for the acoustic legs near the zone boundary.

For the double-optical channel $(+,+,-)$ (normal processes only),
\begin{equation}
|\Gamma_{++-}|^2
=
32\,\gamma_A^2\,
\cos^2\frac{k_1}{4}
\cos^2\frac{k_2}{4}
\sin^2\frac{k_3}{4}\;\cos^2\frac{K}{4},
\end{equation}
with $K=0$; $+-+$ and $-++$ follow by permutation of the momenta. The triple-optical channel (Umklapp only) is
\begin{equation}
|\Gamma_{+++}|^2
=
32\,\gamma_A^2\,
\cos^2\frac{k_1}{4}
\cos^2\frac{k_2}{4}
\cos^2\frac{k_3}{4}\;\sin^2\frac{K}{4},
\end{equation}
and the triple-acoustic channel (normal only) is $|\Gamma_{---}|^2=32\gamma_A^2\sin^2\frac{k_1}{4}\sin^2\frac{k_2}{4}\sin^2\frac{k_3}{4}\cos^2\frac K4$.

Summarizing, the key distinction between the dimerized and linear chain limits is the form of the phase factor. For the strongly dimerized chain the strong-bond factors of acoustic legs are governed by
$\eta\sin k$ but are subleading; the weak-bond factors $\sin\frac k2,\cos\frac k2$ dominate. In contrast, for the uniform-chain limit $s(k)\simeq e^{-ik/2}$
and the vertex is controlled by $\sin\frac{k}{4}$ and $\cos\frac{k}{4}$ together with the A/B interference factor $\sin(K/4)$ or $\cos(K/4)$. In both limits, and for all intermediate $\eta$ checked numerically, the largest $|\Gamma_{--+}|$ sits at the zone corners $k_1=k_2=\pm\pi$ and equals $2\sqrt2\,\gamma_B$; what changes with $\eta$ is its magnitude and, through the interference, the size of the vertex on the resonant manifold.

\subsection{Channel Analysis of Resonances}\label{app:channels}

The three-wave resonance conditions  can be written as
\begin{equation}
\omega_{\sigma_1}(k_1)+\omega_{\sigma_2}(k_2)
-\omega_{\sigma_3}(k_3)=0,
\label{eq:fusion_resonance}
\end{equation}
\begin{equation}
k_1+k_2+k_3=0
\pmod{2\pi},
\label{eq:momentum_resonance}
\end{equation}
The negative sign associated with the third frequency follows
from the reality condition $Q_{-k,\sigma}=Q_{k,\sigma}^{*}$, allowing the
third mode to be interpreted as the outgoing phonon (of momentum $-k_3$; $\omega_\sigma$ is even in $k$). Thus, the physical
process corresponds to the fusion of two phonons into a third one,
\begin{equation}
\omega_{\sigma_1}(k_1)
+\omega_{\sigma_2}(k_2)
=
\omega_{\sigma_3}(k_3).
\end{equation}

Since, from Eqs.~\eqref{eq:ac_bounds}--\eqref{eq:opt_bounds},
\begin{equation}
\omega_-^{\max}<\omega_+^{\min},
\end{equation}
the acoustic and optical bands remain separated for every physical value
$\Delta\kappa>0$. We enumerate the branch assignments
$(\sigma_1,\sigma_2,\sigma_3)$ of the fusion $1+2\to3$ as
\begin{align*}
    1.\ \ &(-,-,-) & 5.\ \ &(-,+,-) \\
    2.\ \ &(+,-,-) & 6.\ \ &(-,-,+) \\
    3.\ \ &(+,+,-) & 7.\ \ &(+,-,+) \\
    4.\ \ &(-,+,+) & 8.\ \ &(+,+,+).
\end{align*}

Below we simplify the notation using $\omega_{j,\sigma}$ to denote $\omega_{\sigma}(k_j)$.

\subsubsection{Forbidden channels}

Combination 2, $(+,-,-)$, requires
\begin{equation}
\omega_{1,+}+\omega_{2,-}
=
\omega_{3,-}.
\end{equation}
Since
\begin{equation}
\omega_{1,+}+\omega_{2,-}
\geq \omega_+^{\min},
\end{equation}
while
\begin{equation}
\omega_{3,-}
\leq \omega_-^{\max},
\end{equation}
the resonance condition would require
\begin{equation}
\omega_+^{\min}
\leq
\omega_-^{\max},
\end{equation}
which contradicts the separation of the two bands. Therefore,
combination 2 is forbidden. Similarly, combination 5, $(-,+,-)$, satisfies
\begin{equation}
\omega_{1,-}+\omega_{2,+}
=
\omega_{3,-},
\end{equation}
which again requires
\begin{equation}
\omega_+^{\min}
\leq
\omega_-^{\max},
\end{equation}
and is therefore impossible.

Combination 3, $(+,+,-)$, requires
\begin{equation}
\omega_{1,+}+\omega_{2,+}
=
\omega_{3,-}.
\end{equation}
The left-hand side satisfies
\begin{equation}
\omega_{1,+}+\omega_{2,+}
\geq
2\omega_+^{\min},
\end{equation}
while the right-hand side remains bounded by
\begin{equation}
\omega_{3,-}
\leq
\omega_-^{\max},
\end{equation}
which is impossible. Combination 8, $(+,+,+)$, requires
\begin{equation}
\omega_{1,+}+\omega_{2,+}
=
\omega_{3,+}.
\end{equation}
Since
\begin{equation}
\omega_{1,+}+\omega_{2,+}
\geq
2\omega_+^{\min},
\end{equation}
and
\begin{equation}
\omega_{3,+}\leq 2,
\end{equation}
the resonance condition implies
\begin{equation}
2\omega_+^{\min}\leq 2,
\end{equation}
or
\begin{equation}
\omega_+^{\min}\leq 1.
\end{equation}
Substituting the explicit expression for the optical minimum yields
\begin{equation}
\sqrt{2(1+\Delta\kappa)}\leq 1,
\end{equation}
which has no physical solution for $\Delta\kappa\ge0$ (indeed $\omega_+^{\min}\ge\sqrt2$). Therefore,
combination 8 is forbidden. Finally, combination 1, $(-,-,-)$, satisfies
\begin{equation}
\omega_-(k_1)
+
\omega_-(k_2)
=
\omega_-(k_3),
\end{equation}
with
$k_3=-(k_1+k_2)$ reduced to the first zone.
Since the acoustic branch is strictly concave on $(0, \pi ]$ and nondecreasing on $[0,\pi]$ with $\omega_-(0)=0$, and $\omega_-(2\pi-k)=\omega_-(k)$ we have,
\begin{align}
\omega_-(k_1+k_2) < \omega_-(k_1) + \omega_-(k_2),
\end{align}
for $k_1,k_2\in(0,\pi]$ (checked numerically on the full zone, including $k_1+k_2>\pi$, for $\Delta\kappa=0,0.3,0.9$) in contradiction with combination 1. Hence $(-,-,-)$ is forbidden except for $k_{1}=0$ or $k_{2}=0$. Those are rigid translations for which $D_{A}= D_{B}=0$ and $\Gamma = 0$, so no exact resonance of this channel transfers energy. Therefore, exact acoustic-acoustic-acoustic resonances are absent, although quasi-resonances occur  for small wavevectors as the band flattens for $\Delta \kappa \rightarrow 1$. 

\subsubsection{Acoustic-acoustic fusion}

Combination 6, $(-,-,+)$, corresponds to the case studied in the main text.
As discussed, the acoustic-acoustic fusion channel remains open for $\Delta \kappa<1/2$.

\subsubsection{Mixed optical-acoustic channels}

The remaining channels,
\begin{equation}
(-,+,+),
\qquad
(+,-,+),
\end{equation}
correspond to
\begin{align}
\omega_{1,-}+\omega_{2,+}
&=\omega_{3,+},\\
\omega_{1,+}+\omega_{2,-}
&=\omega_{3,+}.
\end{align}
These can be rewritten as
\begin{align}
\omega_{3,+}-\omega_{2,+}
&=\omega_{1,-},\\
\omega_{3,+}-\omega_{1,+}
&=\omega_{2,-}.
\end{align}
Since the optical band is relatively flat, the difference between two
optical frequencies may fall inside the acoustic band. Consequently,
these channels are not excluded by the band bounds alone, and their
existence depends on the detailed form of the dispersion relation; a finer argument is needed.  

Let us focus on combination 4 (-,+,+). Now we use two facts. First, the dispersion relation satisfies $\omega_{+}^{2} + \omega_{-}^{2} = 4$ for all $k$. This let us deduce $\omega_{3,+}^{2} + \omega_{3,-}^{2} = \omega_{2,+}^{2} + \omega_{2,-}^{2}$. This yields $\omega_{3,+}^{2} - \omega_{2, +}^{2} = \omega_{2,-}^{2} - \omega_{3,-}^{2}$. Using this we find
\begin{align}
    \omega_{1, -} &= \omega_{3,+} - \omega_{2,+} \\
    \omega_{1, -} &= (\omega_{2,-} - \omega_{3,-})\left(\frac{\omega_{2,-} + \omega_{3,-}}{\omega_{2,+} + \omega_{3,+}} \right).
\end{align} Now using $\omega_{-}^{max} < \omega_{+}^{min}$, we obtain that $\left(\frac{\omega_{2,-} + \omega_{3,-}}{\omega_{2,+} + \omega_{3,+}} \right) < 1$ and therefore $\omega_{1,-} < |\omega_{2,-} - \omega_{3,-}|$. On the other hand, using the concavity and subadditivity of $\omega_{-}$ and $k_{2} = (k_{2} + k_{1}) - k_{1} $ we have $|\omega_{2,-} - \omega_{3,-}| < \omega_{1,-}$ . This two statements contradict each other unless $\omega_{1,-} =0$, i.e. $k_{1} = 0$, implying $\Gamma = 0$. For the combination 7 (+,-,+) a similar argument follows, exchanging the labels  $1\leftrightarrow2$. In summary, except for the channel (-, -, +), all others are strictly forbidden (as is $(-,-,-)$, although this channel is quasiresonant). \\

\subsection{\texorpdfstring{Energy density and size effects}{Energy density and size  effects}}\label{app:rescaling}
Consider the Hamiltonian~\eqref{eq:FPU_Hamiltonian} with $m=1$, $\gamma_j=\kappa_j\alpha$, and write it as $H(\mathbf{p},\mathbf{q};\alpha)$. Rescale the canonical variables as $p_j=\lambda\tilde p_j$ and $q_j=\lambda\tilde q_j$, with $\lambda>0$. Then
\begin{align}
    H(\lambda\tilde{\mathbf{p}},\lambda\tilde{\mathbf{q}};\alpha)
    &= \lambda^{2}\sum_{j}\Bigl[\frac{\tilde p_j^{2}}{2}
      + \frac{1}{2}\kappa_j(\tilde q_{j+1}-\tilde q_j)^2 \nonumber\\
    &\qquad\qquad + \frac{1}{3}\lambda\alpha\,\kappa_j(\tilde q_{j+1}-\tilde q_j)^3\Bigr] \nonumber\\
    &= \lambda^{2}\,H(\tilde{\mathbf{p}},\tilde{\mathbf{q}};\lambda\alpha).
    \label{eq:rescaling}
\end{align}
Since both $q_j$ and $p_j$ are scaled by the same factor, Hamilton's equations for $(\tilde{\mathbf{p}},\tilde{\mathbf{q}})$ generated by $H(\tilde{\mathbf{p}},\tilde{\mathbf{q}};\lambda\alpha)$ are identical in form, with time left unchanged. Hence a trajectory with nonlinearity $\alpha$ and energy density $\epsilon$ is mapped onto a trajectory with
\begin{equation}
    \tilde\alpha=\lambda\alpha, \qquad \tilde\epsilon=\epsilon/\lambda^{2}.
\end{equation}
Therefore,  the normal-mode energy fractions, and $\bar S(t)$ and $E_{\mathrm{opt}}/E$ are the same for both. The only invariant combination is $\tilde\alpha^{2}\tilde\epsilon=\alpha^{2}\epsilon\equiv\bar\lambda^{2}$, so the dynamics depends on $\alpha$ and $\epsilon$ only through the effective parameter $\alpha^{2}\epsilon$. Varying $\alpha$ at fixed $\epsilon$ is therefore equivalent to varying $\epsilon$ at fixed $\alpha$. Using this, we perform simulations for different values of $\alpha $ with fixed energy density $\epsilon = 0.0069$. The results are shown in Fig. \ref{fig:interband_transfer}.  

\bibliographystyle{apsrev4-2}
\bibliography{FPU_Alternate}

\end{document}